\documentclass[fleqn,usenatbib]{mnras}

\usepackage{newtxtext,newtxmath}

\usepackage[T1]{fontenc}
\usepackage{ae,aecompl}

\usepackage{graphicx}    
\usepackage{amsmath}    
\usepackage{bm}
\usepackage{natbib}
\usepackage{color,units}
\usepackage{float}
\usepackage{caption}
\usepackage{aas_macros}
\usepackage{hyperref}
\usepackage{widetext}
\usepackage{caption}
\usepackage{subcaption}
\usepackage[normalem]{ulem}

\usepackage{listings}
\usepackage{xcolor}
 
\definecolor{codegreen}{rgb}{0,0.6,0}
\definecolor{codegray}{rgb}{0.5,0.5,0.5}
\definecolor{codepurple}{rgb}{0.58,0,0.82}
\definecolor{backcolour}{rgb}{0.95,0.95,0.95}
 
\lstdefinestyle{mystyle}{
    backgroundcolor=\color{backcolour},   
    commentstyle=\color{codegreen},
    keywordstyle=\color{magenta},
    numberstyle=\tiny\color{codegray},
    stringstyle=\color{codepurple},
    basicstyle=\ttfamily\footnotesize,
    breakatwhitespace=false,         
    breaklines=true,                 
    captionpos=b,                    
    keepspaces=true,                 
    numbers=none,                    
    numbersep=5pt,                  
    showspaces=false,                
    showstringspaces=false,
    showtabs=false,                  
    tabsize=2
}
 
\newcommand{\citeg}[1]{\citep[e.g.,][]{#1}}
\newcommand{\getdata}{{\tt get\_data}}
\newcommand{\transient}{{\tt Transient}}

\newcommand{\result}{{\tt result}}

\newcommand{\likelihood}{{\tt likelihood}}
\newcommand{\lasair}{{\tt LASAIR}}

\newcommand{\jointlikelihood}{{\tt joint\_likelihood}}
\newcommand{\prior}{{\tt prior}}

\newcommand{\surrogate}{\texttt{redback\_surrogate}}
\newcommand{\mosfit}{\texttt{MOSFiT}}
\newcommand{\constraints}{{\tt prior\_constraints}}
\newcommand{\simulation}{{\tt simulation}}
\newcommand{\transientmodels}{{\tt transient\_models}}
\newcommand{\redbackurl}{\url{https://github.com/nikhil-sarin/redback}}
\newcommand{\redbackdocs}{\url{https://redback.readthedocs.io/en/latest/}}

\newcommand{\swift}{\textit{Swift}}
\newcommand{\bilby}{{\sc Bilby}}
\newcommand{\pymultinest}{{\sc pymultinest}}
\newcommand{\nestle}{{\sc nestle}}
\newcommand{\dynesty}{{\sc dynesty}}
\newcommand{\redback}{{\sc Redback}}
\newcommand{\program}[1]{\textsc{#1}}

\newcommand{\python}{{\texttt{python}}}

\newcommand{\UniMelb}{School of Physics, University of Melbourne, Parkville, Victoria 3010, Australia}
\newcommand{\OzGravUniMelb}{OzGrav: The ARC Centre of Excellence for Gravitational Wave Discovery, University of Melbourne, Parkville, Victoria 3010, Australia}
\newcommand{\OkcPhys}{The Oskar Klein Centre, Department of Physics, Stockholm University, AlbaNova, SE-106 91 Stockholm, Sweden}
\newcommand{\Okcastro}{The Oskar Klein Centre, Department of Astronomy, Stockholm University, AlbaNova, SE-106 91 Stockholm, Sweden}

\title[\sc{Redback}]{{\sc Redback}: A Bayesian inference software package for electromagnetic transients}
\author[N.Sarin et al.]{Nikhil Sarin,$^{1,2}$\thanks{E-mail: nikhil.sarin@su.se},
Moritz H\"{u}bner$^{3,4}$,
Conor M. B. Omand$^{5}$,
Christian N. Setzer$^{2}$,  
Steve Schulze$^2$, \newauthor
Naresh Adhikari$^{6}$, 
Ana Sagu{\'e}s-Carracedo$^{2}$,
Shanika Galaudage$^{7,8}$, 
Wendy F. Wallace$^{9, 1}$,  
Gavin P. Lamb$^{10}$, \newauthor
and En-Tzu Lin$^{11}$
\\
$^{1}$Nordita,  Stockholm University and KTH Royal Institute of Technology
Hannes Alfvéns väg 12, SE-106 91 Stockholm, Sweden\\
$^{2}$\OkcPhys\\
$^{3}$\UniMelb\\
$^{4}$\OzGravUniMelb\\
$^{5}$\Okcastro\\
$^{6}$Leonard E. Parker Center for Gravitation, Cosmology, and Astrophysics, University of Wisconsin-Milwaukee, Milwaukee, WI 53201, USA \\
$^{7}$Université Côte d'Azur, Observatoire de la Côte d'Azur, CNRS, Laboratoire Lagrange, Bd de l'Observatoire, F-06304 Nice, France\\
$^{8}$Université Côte d'Azur, Observatoire de la Côte d'Azur, CNRS, Artemis, Bd de l'Observatoire, F-06304 Nice, France\\
$^{9}$Department of Physics, University of Bath, Claverton Down, Bath, BA2 7AY, UK \\
$^{10}$ Astrophysics Research Institute, Liverpool John Moores University, IC2 Liverpool Science Park, 146 Brownlow Hill, Liverpool, L3 5RF, UK \\
$^{11}$Institute of Astronomy, National Tsing Hua University, Hsinchu 30013, Taiwan}

\date{Accepted XXX. Received YYY; in original form ZZZ}

\pubyear{2023}

\begin{document}
\label{firstpage}
\pagerange{\pageref{firstpage}--\pageref{lastpage}}
\maketitle

\begin{abstract}
Fulfilling the rich promise of rapid advances in time-domain astronomy is only possible through confronting our observations with physical models and extracting the parameters that best describe what we see. 
Here, we introduce {\sc Redback}; a Bayesian inference software package for electromagnetic transients. {\sc Redback} provides an object-orientated {\sc python} interface to over 12 different samplers and over 100 different models for kilonovae, supernovae, gamma-ray burst afterglows, tidal disruption events, engine-driven transients among other explosive transients. The models range in complexity from simple analytical and semi-analytical models to surrogates built upon numerical simulations accelerated via machine learning. {\sc Redback} also provides a simple interface for downloading and processing data from various catalogs such as \textit{Swift} and Fink. The software can also serve as an engine to simulate transients for telescopes such as the Zwicky Transient Facility and Vera Rubin with realistic cadences, limiting magnitudes, and sky-coverage or a hypothetical user-constructed survey or a generic transient for target-of-opportunity observations with different telescopes. As a demonstration of its capabilities, we show how {\sc Redback} can be used to jointly fit the spectrum and photometry of a kilonova, enabling a more powerful, holistic probe into the properties of a transient.
We also showcase general examples of how {\sc Redback} can be used as a tool to simulate transients for realistic surveys, fit models to real, simulated, or private data, multi-messenger inference with gravitational waves, and serve as an end-to-end software toolkit for parameter estimation and interpreting the nature of electromagnetic transients.
\end{abstract}
\section{Introduction}\label{sec:intro}
Rapid advances in electromagnetic telescope sensitivity and survey capabilities are revolutionising transient astronomy.
However, to realise the full promise of the rich and large photometric and spectroscopic data sets, we need a robust toolkit for simulating what we expect to see, building and exploring our models and fitting the observations. Such advancements can enable us to ultimately learn the physics that drives these transients, optimise our survey strategies and instruments, and gain insights into the lives and afterlives of stars and the evolution of our Universe.
Such a tool must also be modular and open-source, easily adaptable to an individual user's needs and efficiently maintained and upgraded.

Several iterations of open-source software have served important roles in improving our understanding of transients. For example, \texttt{MOSFiT}~\citep{Guillochon2018}, a modular package that has been used for parameter estimation of several electromagnetic transients such as tidal disruption events~\citep{Mockler+19}, superluminous supernovae~\citep{Nicholl2017}, and kilonovae~\citep{Villar2017}. The \texttt{SNCosmo}~\citep{Barbary2022}, and \texttt{SNANA}~\citep{Kessler2009} software suites that are readily used to fit Type Ia supernovae to enable cosmological analyses~\citeg{Vincenzi2024} or study the detectable rates of supernovae for different survey designs~\citeg{Bom2024}. \texttt{3ML}~\citep{Vianello2015}, which provides a cohesive framework utilising existing instrument-specific software to best capture how the data is generated and perform detailed modelling of gamma-ray bursts across data from multiple instruments~\citeg{Klinger2024}. \texttt{Haffet}~\citep{Yang2023}, which enables data-driven reconstruction of supernova bolometric luminosity from multi-band photometry enabling a more direct probe to study the explosion properties~\citeg{Dong2023}. \texttt{NMMA}~\citep{Pang2022}, that provides machine-learning based ``surrogates`` to radiative transfer simulations of kilonovae, enabling inference with kilonova models that include the most physics. The \texttt{NMMA} package also provides an interface for jointly analysing electromagnetic and gravitational-wave data such as for the first gravitational-wave observation of a binary neutron star merger, GW170817~\citep{abbott17_gw170817_multimessenger}, enabling strong constraints on the behaviour of nuclear matter~\citep{Koehn2024}.

Although these software packages have driven significant progress in electromagnetic transient astronomy, several limitations must be addressed to take full advantage of the currently available and forthcoming electromagnetic data. For example, models for explosive transients are under constant development and often make several underlying assumptions. However, these packages above are limited to a small library of implemented models and inflexible interfaces to change or add new models. This prevents detailed studies into modelling systematics or the use of the best models for any given transient. To truly leverage the data and maximally extract insights into these transients, open-source packages must come equipped with a large variety of built-in models and are routinely updated to capture the best theory has to offer. Ideally, such packages also provide a simplified interface to enable end-users to drop-in replacements or modify features for inbuilt models or with minimal interaction with the source code. The last point is pertinent as this could help remove the burden on development teams to keep pace and implement developments from transient modelling, particularly in the scenario where key developers leave the field, as has been the case of some of the above packages.

Similarly, there are constant improvements to Bayesian inference techniques that are not captured by several of these packages above as they typically use at most one sampling package such as \program{emcee}~\citep{Foreman-Mackey2015}. It is worth noting that some of these packages are also not Bayesian, failing to provide robust estimation of the uncertainty in estimating parameters from any fit. The lack of multiple implemented packages prevents cross-sampler validation (a valuable tool to determine if results are robust) or leveraging the benefits of different samplers, such as evidence calculation for Bayesian model selection. Other sampling algorithms can also be better tuned for specific transient problems (dramatically improving sampling wall-clock time), allowing for the use of the best tools for the task at hand. Similarly, there are also several practical benefits to having access to multiple different sampling algorithms, such as a better ability to capture multi-modal posterior distributions or parallelisation.

A critical validation step in any inference workflow is to test how models perform across the parameter space and tests with complete end-to-end analyses, i.e., from simulation to fitting workflow. While the above packages have been tested in various ways, they do not all provide a cohesive framework to both simulate model outputs (for a variety of different formats such as flux density or magnitudes or bolometric luminosity) and realistic observations (for real surveys or target-of-opportunity observations) and fit them. This is a limitation of many of these packages, as we can only truly determine bias in parameter estimation by performing simulations with the same tools we use to fit and control the data generation process. Properly capturing the data-generation process is also instrumental for accurate transient analyses. The bulk of the above packages can only work with the simple assumption of a Gaussian likelihood (i.e., the noise distribution is Gaussian around the true input model). This simple noise assumption is known to be incorrect for the majority of current and projected future observations and will undoubtedly cause problems as we continue to observe each transient more frequently and with higher fidelity.

Higher fidelity and more extensive observations of transients also open up another challenge to maximally leverage our data: astronomical events are now readily observed in multiple ways. An example already described above was the multi-messenger gravitational-wave discovery of the binary neutron star merger GW170817~\citep{abbott17_gw170817_multimessenger}, which had an afterglow observed across the electromagnetic spectrum~\citep{Hallinan2017, Alexander2018, Fong2019, Lamb2019_170817}, a kilonova from near-infrared to ultra-violet~\citep{Pian2017, Smartt2017, Villar2017}, very-long baseline interferometry (VLBI)~\citep{Mooley2018}, and gravitational-wave data~\citep{Abbott2017}. While packages such as \texttt{NMMA} can perform a joint analysis of the gravitational wave and electromagnetic photometry, they fail to include the spectrum or VLBI data. Although these constraints could be folded through after the photometric and gravitational-wave analysis, you then lose the significant benefits offered by the full Bayesian framework~\citep{Ryan2023, Gianfagna2023}. The opportunity to jointly fit the spectrum and photometry also provides a holistic look into the properties of the transient, where the photometric and spectroscopic analyses can often tell a contradictory story. Moreover, a flexible framework for combining datasets could also enable Bayesian hierarchical modelling, a powerful technique to uncover the properties of a population.

Here, we introduce \redback{}, an open-source, end-to-end Bayesian Inference software package for simulating and fitting electromagnetic transients.
\redback{} provides an object-orientated \python{} interface to over 12 sampling software and over 100 models for several different electromagnetic transients. Furthermore, \redback{} provides a simplified interface to download data for multiple transients from various catalogues, handling processing to a homogeneous format, removing the burden from end-users to fully understand the peculiarities of different data sources. For all models implemented in \redback{} or user-provided models, end-users of \redback{} can simulate transients for actual surveys such as the Large Synoptic Survey of Space and Time (LSST)~\citep{Ivezic2019} and Zwicky Transient Facility (ZTF)~\citep{ztf_paper},  or a custom survey, alongside target-of-opportunity observations for any collection of observatories/telescopes. Users can fit this simulated, private, or publicly available data through Bayesian inference alongside combinations of different data types such as VLBI data, gravitational-wave data, and a transient spectrum and photometry. \redback{} is also built on modern \python{}, with many adopted practices to aid continual development, continuous integration, and an extensive library of unit tests and examples which ensure that the primary features of \redback{} remain stable through future development.

\redback{} provides several advantages over other software packages and mitigates the aforementioned issues:
\begin{itemize}
\item An extensive library of inbuilt models and a simple interface for users to add their own. Several models implemented in \redback{} are direct improvements to previous models or model transients one can not model in other packages.
\item An engine to simulate realistic transients for surveys and target of opportunity observations and perform inference, i.e., a tool to validate an entire inference workflow or optimise surveys.
\item A tool to access and process photometric data alongside auxiliary data such as sky position from many publicly available catalogs and brokers.
\item A modular and flexible interface, users can swap likelihoods, models, and plotting without ever modifying the source code. Alternatively, change how existing aspects of \redback{} function by passing their own function to existing functional modules.
\item Simplified interface (replace a string) to over 12 different open-source samplers, enabling cross-sampler validation or use of samplers better tuned for transient inference or have additional capabilities such as multiprocessing.
\item Modern \python{} software development practices, including continuous integration and unit-testing to ensure core software features remain functional, even if core developers leave the field.
\end{itemize}

This paper is intended to describe the capabilities and mark the version 1.0 release of \redback{}. \redback{} is installable via \texttt{pip} and available at \redbackurl{}. We note that \redback{} has been open-source and distributed under the GPL licence since March 2022. Earlier versions of \redback{} have already been used in previous publications, which we refer the interested reader to see some of the use cases for \redback{}~\citep{Sarin2020_collapsing, Sarin2020_radiative, Sarin2021_cdf, Sarin2022_blt, Sarin2022_mag, Sarin2022_nemo, Schulze2023, Levan2023, SarinMetzger23, OmandSarin23, Rosswog2024}.

This paper is structured as follows: In Sec.~\ref{sec:design} we describe the design objectives of \redback{}, how the different parts of the software interact, and the $3$ typical workflows we expect \redback{} to be used for.
In Sec.~\ref{sec:overview}, we describe the different functional modules in \redback{} and how they are used in various workflows. In Sec.~\ref{sec:joint_spectrum_photometry}, we showcase a new scientific analysis enabled by \redback{}; the joint fitting of the spectrum and photometry of a kilonova. In Sec.~\ref{sec:future} we briefly describe features in \redback{} that will be added in future releases and conclude in Sec.~\ref{sec:conclusion}. In the Appendix, we showcase the general interface with detailed code snippets of how to use \redback{}, alongside more detailed examples. In particular,  in Sec.~\ref{sec:interface}, we show how to download and process data, set up the inference workflow, several plotting methods and simulate transients. This basic interface is followed by more detailed examples in Sec.~\ref{sec:multimessenger} and \ref{sec:examples} where we demonstrate the different capabilities of \redback{}. In particular, we first show how \redback{} can be used to jointly analyse a multi-messenger binary neutron star signal with X-ray and gravitational-wave data and then to fit different types of real electromagnetic transients.
\begin{figure*}
    \centering
    \includegraphics[width=1.00\textwidth]{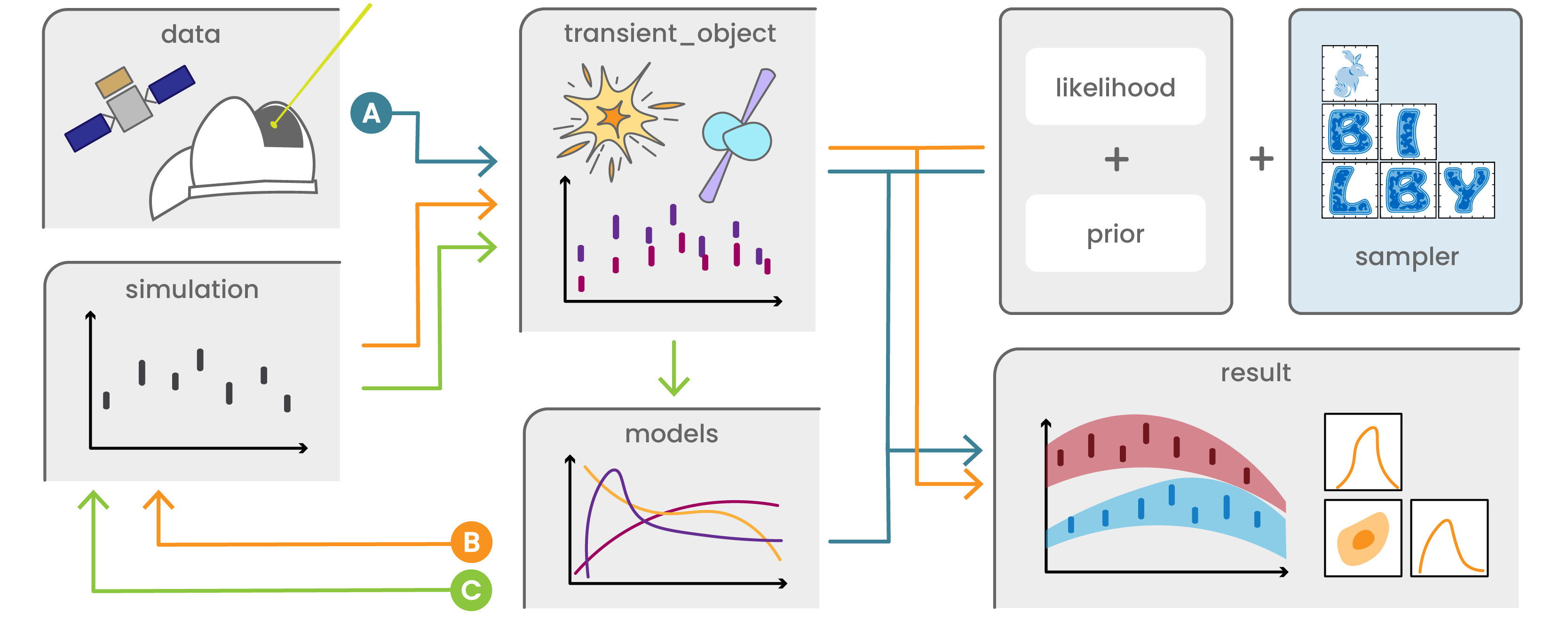}
    \caption{Flowchart showcasing the different subpackages, modules, and classes} of \redback{} and how they interact for different workflows.
    \label{fig:flowchart}
\end{figure*}

\section{Design and Implementation}\label{sec:design}
A core design objective for \redback{} is to be truly modular, with the flexibility to adapt to the different requirements/preferences of end-users and for users to use different parts of the software without requiring additional overhead or modifying the source code. 

Second, \redback{} must be flexible to both serve as a workhorse in expert workflows in transient astronomy and as an accessible tool for newcomers to the field. In particular, while advanced users can be expected to modify or interact more directly with various aspects of the \redback{} software, novice users must be able to use all aspects of \redback{} from data collection, simulation, and fitting with just a few lines of code.

Third, where possible, we also aim to leverage other open-source software to reduce the burden on core developers of \redback{} and better keep pace with developments in other areas. For example, we are tightly integrated with the \bilby{} framework for sampling. This provides a simplified interface for end users to multiple open-source samplers and access to a large and active development team that maintains \bilby{} and different sampling packages.

To address these design objects, all primary functional modules of \redback{} are built as \python{} classes or functions. These functional modules can be readily modified by end-users via keyword arguments or replaced entirely within a workflow with other functional modules implemented in \redback{} or something the user provides. This modularity extends to primary modules described below and to more practical features such as plotting or where \redback{} outputs are stored.

This modular interface addresses some critical limitations with previous packages described in the introduction. For example, all \redback{} models are implemented as callable \python{} functions. These models also have minimal dependencies and do not depend on other aspects of the software, enabling users to evaluate a model as they would with any other \python{} function. Moreover, all models in \redback{} can produce different outputs, e.g., bolometric luminosity, a spectrum, a flux, a magnitude, a flux density, or auxiliary information such as the photospheric velocity. For many \redback{} models, users can also change critical assumptions of the model, such as the spectral energy distribution, by passing in a different keyword argument. This allows end-users to generate model outputs for any arbitrary input, better understand the effects of different parameters, or change modelling assumptions without modifying the source code. It also facilitates holistic studies by fully considering various observations, e.g., spectrum and photometry. End-users can also replace a model with their own \python{} function, keeping intact all of the other functionality of \redback{}, making it significantly easier to use new and improved models with \redback{}.

Advanced users can also change the likelihood, i.e., their assumptions about the data-generating process or the prior distribution on model parameters with different implementations in \redback{} or their implementation, again, without needing to change the source code. This enables advanced users to adapt their fitting or simulation workflows for more sophisticated analyses. At the same time, such choices are made by default for novice users, who can perform such tasks with minimal domain expertise. Moreover, different sampling algorithms and software packages can be used with minimal effort by simply changing the string referring to a sampler. This addresses the limitations of previous packages with inflexible interfaces for users to change assumptions about how the data is processed or generated and leverage the best samplers for the respective task.

In Fig.\ref{fig:flowchart} we show the different functional modules of \redback{}, implemented as either a \python{} subpackage, a \python{} module, a \python{} class or as a \python{} function and how they interact for the $3$ most common workflows we expect this software to be used for. 

\begin{itemize}
    \item[] {\bf A: Fitting a real transient.} We anticipate that one of the most common use cases for \redback{} will be fitting data of a real astrophysical transient. This workflow typically involves getting data from one of the catalogues using the {\tt get\_data} subpackage in \redback{}, or users can provide private data. The user will then use this data to create a specific \transient{} class object from the transient subpackage, which loads the data in a homogeneous format and can be used for plotting the data or additional processing, such as converting flux data to luminosity. The user then passes this \transient{} class object along with a string referring to a model from \transientmodels{} subpackage or their own \python-wrapped model, an instance from the prior {class}, an instance of the likelihood class and a string referring to a sampler that is available in \bilby{}~\citep{bilby1,bilby2}. This will perform fitting through Bayesian inference and obtain a \result{} class object. The \result{} object contains the posterior, and other properties such as the Bayesian evidence. This \result{} class can also be used to make plots such as the fitted lightcurve, the corner plot, or the cumulative distribution function of all parameters, which are internally handled by a separate plotting module. We emphasise that for the novice user, choices like the likelihood, prior, sampler choice and plot aesthetics are made by default, but more advanced users can change these as they desire. 
    \item[] {\bf B: Fitting a simulated transient.} Many users will also fit simulated data to verify the inference workflow or predict constraints from mock observations. In this workflow, the user would start with a model from \transientmodels{} or supply their own model. Then, use the \simulation{} module to create synthetic data. After the creation of this simulated data, the workflow for fitting is the same as workflow A. 
    \item[] {\bf C: Simulating a transient or a population of transients.} Users may also wish to create a population of transients, for example, to understand how many afterglows LSST will see in a year or to understand the selection effects of surveys. These workflows require choosing a model from \transientmodels{} or supplying a \python-wrapped model and passing this to the \simulation{} module alongside a \prior{} object which describes the distribution of each parameter in the model that constitutes the population. Complex prior constraints can be placed on this population through the use of \constraints{} functional module.
\end{itemize}

The above briefly describes how different aspects in \redback{} interact for different workflows. We now give a general overview of the \redback{} software and describe each functional module's capabilities in detail and how it can be modified. 
\section{Software Package Overview}\label{sec:overview}
\redback{} is built predominantly on a class structure and almost every aspect of the software exists as an independent \python{} class. Here we describe each of these different functional modules and their primary functionality. 
We stress that these modules are standalone and can be used independently to adapt to different needs and workflows, or modified via keyword arguments or replaced to provide additional functionality. 
\subsection{Data interface}
\redback{} provides an interface to download and process data from multiple catalogues through the \getdata{} subpackage. In particular, this includes the flux, flux density or the photon arrival time data for gamma-ray bursts (GRBs) detected by the \textit{Neil Gehrels Swift Observatory} available at Swift Data Centre~\citep{Evans2010}, the magnitude or flux density data of transients from ZTF from \lasair{}~\citep{Smith_2019} or Fink~\citep{Moller2021} which in the future are expected to also host transient lightcurves from LSST~\citep{Ivezic2019}, the archival GRB data from BATSE~\citep{Fishman1994} and compilation of optical transient lightcurves available at the Open Access Catalog (OAC)~\citep{Guillochon2017}.

For each of the above catalogues, the \getdata{} module provides a one line interface to download and process the data into a \texttt{pandas} data frame and save it as a human-readable file in an appropriate location to integrate with the rest of \redback{}. 
This module also attempts to find additional metadata such as the redshift of the transient, the GRB photon index, T$90$ among other properties and process the data to add additional attributes such as the integrated flux, flux density and their respective errors.

As the data is stored as human-readable file and readable as a \texttt{pandas} data frame, the user can easily add additional private data or verify and modify any erroneous data. The \getdata{} can be used independently of all other parts of \redback{} and may be used to simply process a large quantity of transient data from public archives. 
\subsection{Transient classes}
The primary unifying module of \redback{} are the \transient{} classes that are available through the Transient subpackage. These are separated into two main types, a generic transient class which is applicable for any type of transient and an optical transient class. These classes serve as parent classes for five other classes; \texttt{prompt}, \texttt{afterglow}, \texttt{kilonova}, \texttt{supernova}, and \texttt{tde} which provide a more seamless interface for the specific type of transient, any additional processing such as converting the flux data to a luminosity and modify some default behaviour, such as labels for plotting, where plots are saved etc. 
We note that the \texttt{afterglow} class is further split into a short and long GRB class but these are functionally equivalent and only differ in locations of metadata. 

For all transient classes, we provide one-line class methods to load the data from different catalogs obtained via the \getdata{} module, or from the \texttt{simulation} module (described in Sec \ref{sec:simulation}). The \transient{} objects can also be initialised independently of any class method by specifying the observed properties. In Sec.~\ref{sec:creatingtransients}, we show how to initialise these \transient{} objects for different workflows. 

All transient objects also have two important attributes; \texttt{data\_mode} and \texttt{use\_phase\_model}. The former is an attribute which dictates what \redback{} assumes to be the mode of data for the transient e.g., \texttt{magnitude} for magnitude data, while the latter is a Boolean switch which dictates whether the transient has observed times in reference to a known start time (as usually the case for an afterglow) or is in Modified Julian Days (MJD) without a reference (as usually the case for most other transients). Again, these attributes affect choices such as the labels for plotting and where plots are saved but also in some cases the default likelihood used by \redback{}.
\subsection{Models}
While the most desirable method would be to confront observations with the best models that include the most physics (typically hydrodynamical and radiative transfer simulations), such models are not tractable for fitting given the demanding computational requirements of Bayesian inference (each model must be evaluated over a range of parameters at least $\mathcal{O}(10^4)$ times to fit a typical transient). To be tractable for inference, all models in \redback{} are either analytical, semi-analytical, or surrogates built with machine learning from numerical simulations. 
The latter are provided by another standalone software package~\surrogate{} that is available independently but we consider as part of the \redback{} software stack. Here, we describe the models for various different transients available in \redback{} and how they can be modified. 

To remain true to our driving aim of modularity, all models are callable \texttt{python} functions and can be called on an arbitrary set of values with minimal dependencies. 
These functions can all be easily modified through the use of dependency injection (described in Sec \ref{sec:dep_inj}) without needing modify the \redback{} source code or replaced entirely within the rest of the workflow with a user-provided model. Most \redback{} models can provide outputs in different formats e.g., luminosity, integrated flux, magnitude or flux density enabling them to be used to fit any type of data format. For magnitude and integrated flux data, \redback{} will integrate the spectrum and calculate the band pass magnitude/flux. This behaviour could be easily modified to use a flux density to magnitude conversion to further alleviate computational demands. We note that this behaviour is enabled by default for afterglow models where the effect of assuming a flux density to magnitude conversion as opposed to integrating a band pass is minimal.
\subsubsection{User-defined models and dependency injections}\label{sec:dep_inj}
As alluded to above, \redback{} is built on a flexible interface which allows the user to use their model with all other aspects of \redback{}. The only requirement is that the user-defined model is a \python{} function with the first input being the time of observations and the output being the desired output e.g., \texttt{flux\_density} if the user wants to fit \texttt{flux\_density} data. Once written, this \python{} function can be passed to different modules of \redback{}, either to simulate data or to fit some observations. This workflow also enables users to combine \redback{} models, replacing each of the individual models with their own model or a different \redback{} model.

Many \redback{} models use additional keyword arguments to dictate the precise physics of the model. Some keyword arguments are Boolean switches to turn on/off certain physics, but others require a more complex object. This pattern is often referred to as \textit{dependency injection}, which allows us to build a more flexible interface.
We implemented the dependency injection pattern to handle features such as the spectral energy distribution, or the conversion from inspiral parameters to kilonova parameters, photosphere, or the cosmology used to associate a redshift to a luminosity distance. By default, every model has these choices set internally but users can make changes to the model by simply using a different object as a keyword argument which could either be an instance of a \redback{} class or a class they write themselves. 
Through these model modifications and dependency injections, many \redback{} models can be extended and have their physics changed without ever modifying the source code, alleviating the burden on the end-user to make a change to a model. However, as the interface is modular, a \redback{} model can also just be replaced entirely.
\subsubsection{Specific Transient models}
\paragraph*{Broadband GRB afterglow:}

Gamma-ray bursts are typically followed by lower energy broadband emission referred to as afterglow~\citeg{Sari1998}. The broad consensus is that the afterglow is a product of the relativistic jet interacting with the ambient interstellar medium, an interaction that produces synchrotron emission. 
However, there are several aspects of afterglow models that are ill-understood, such as the jet structure, i.e., the distribution of energy as a function of angle, or the role of reverse shocks, or additional emission components, or energy injection. 

\redback{} provides an interface to several different afterglow models. For example, the different jet-structure models implemented in \texttt{afterglowpy}~\citep{Ryan2020}, and implementations of several other physical models described in the literature~\citep{Sari1998, Sari1999, Gottlieb2018, Lamb2020, Lamb2021}. 
For each of the models, users can make additional modifications to the physics such as the inclusion of jet spreading, inverse Compton emission, and energy injection or more specific settings such as the resolution of the integration scheme. 
For other models, users can choose the exact jet-structure profile, whether the interstellar medium is at a constant density or a wind-like medium etc., and whether the shock is refreshed. 
All these modifications are handled through additional optional keyword arguments in the \texttt{python} function which allows the advanced users to make changes as they wish while more novice users can avoid having to make these decisions. 

Alongside these physically-motivated models we also include some purely phenomenological broken power-law models with different degrees of components. In total, \redback{} includes $21$ physically-motivated afterglow models in addition to the five phenomenological models, providing coverage of the different physical assumptions involved in afterglow modelling and to test the robustness of inferred models across different modelling assumptions.
\paragraph*{Broadband kilonova afterglow:}
Similar to a GRB afterglow, there also exists an expectation for synchrotron emission when the slower moving kilonova ejecta interacts with the ambient interstellar medium. However, unlike models for the GRB afterglow, where we are aided by decades of observations, there are currently no confident detections of a kilonova afterglow. 
Nevertheless, we provide an interface to several different kilonova afterglow models described previously in the literature~\citep{Nakar2011, Sarin2022_mag} and make modifications to some of GRB afterglow models described above to be more suited for a kilonova afterglow~\citeg{Gottlieb2018, Ryan2020}. In the future, we will add kilonova afterglow models more representative of the ejecta distribution we see in numerical simulations~\citep{Kathirgamaraju2019, Nedora2021}.
\paragraph*{Kilonovae:}
The revolutionary observations of AT2017gfo~\citeg{abbott17_gw170817_multimessenger, Abbott2017, Arcavi2017, Kasen2017, Coulter2017, Villar2017} provided definitive evidence of a thermal transient powered by r-process nucleosynthesis. However, despite the extensive observations and significant theoretical model development, many aspects of kilonovae remain uncertain. In \redback{} we provide implementations of $18$ different kilonova models which range in complexity and implemented physics. Many aspects of these models such as the distribution of ejecta mass or the recipe to relate the binary neutron star (BNS) or neutron star black hole (NSBH) parameters to kilonova parameters can be changed through the use of dependency injection (described in Sec. \ref{sec:dep_inj}). 

The simplest kilonova model implemented in \redback{} is a one component kilonova model~\citep{Villar2017}. 
Although minimal in parameters and quick to evaluate and therefore fit to observations, this model has already been shown to be unsuccessful in explaining multiple aspects of kilonovae observations. 
To address the inability of such a simple model to explain observations, we also provide implementations of two and three component kilonova models following~\citet{Villar2017} and implementations of \texttt{MOSFiT} kilonova models~\citep{Cowperthwaite2017, Villar2017}. 
These models all effectively ignore the dynamics of the ejecta, assuming the entire ejecta component is moving at one velocity, an assumption that is likely incorrect. We therefore also provide models where the ejecta is distributed into shells which expand homologously, similar in spirit to the model presented in \citet{Metzger2010} and \citet{Metzger2019}. Alongside this, we also provide an interface to the heating-rate kilonova models~\citep{Korobkin2012, Hotokezaka2020, Dorsman2023}, which allow the user to describe the velocity and opacity distribution themselves. 

The above models all have parameters that describe the kilonova ejecta properties itself i.e., the mass and velocity of the ejecta. However, it has become increasingly common for kilonova models to be built upon the BNS or NSBH parameters which are then related to the ejecta parameters with a series of recipes from numerical relativity simulations. We provide several implementations of these models including models for BNS and NSBH, including for example the BNS model implemented in \texttt{MOSFiT}~\citep{Nicholl2021} which includes additional physics such as shock cooling to describe the early optical lightcurve~\citep{Piro2018}, or implementations of models presented in \citet{Coughlin2019}. 

While the above models are all semi-analytical, we also provide $3$ models that are machine-learning surrogates to numerical simulations. These surrogates are provided in the optional package \surrogate{} (described in more detail below), and are implementations of surrogates built in \texttt{KilonovaNet}~\citep{Lukosiute2022}. 
In the future, we will continue to add more kilonovae models~\citep{Banerjee2020, Korobkin2021} and allow greater flexibility to existing models such as changing the calculation of the thermalisation efficiency.
\paragraph*{Supernovae:}
\redback{} contains many supernova models of varying levels of complexity. Most of the models have both a bolometric implementation and an implementation for multi-band photometry. This setup allows the user to fit bolometric luminosity, magnitude, integrated flux, or flux density data. 
Similar to \mosfit{} where physics such as the interaction process, photosphere, and spectral energy distribution (SED) can be swapped, \redback{} supernovae models can do the same since they are implemented using dependency injection. For all models, these aspects are chosen by default corresponding to the physics implemented but can be swapped without modifying the source code for a different module to capture different physics. 

The simplest model, such the exponential-power law model, is purely phenomenological and built upon no physics in terms of luminosity but assumes a diffusive photosphere with a temperature floor, and a blackbody SED. Other models are more physically motivated such as several variations of the Arnett model \citep{Arnett1980, Arnett1982} for $^{56}$Ni-powered supernovae including a version which also incorporates shock cooling, a version that incorporates line absorption for modeling Type Ia supernovae, and a version which incorporates synchrotron emission for modeling type Ic supernovae. Then there are models for circumstellar (CSM) interaction powered supernovae~\citep{Chatzopoulos2013, Villar2017_csm, Jiang2020} as well as a mix of CSM and $^{56}$Ni power. We also include other models similar to those available in \mosfit{}, such as the \texttt{basic magnetar}, \texttt{slsn}, and \texttt{magnetar+nickel} models~\citep{Nicholl2017, Guillochon2018}, as well as new models which include non-vacuum dipole spin-down~\citep{Lasky2017} and ejecta acceleration from the pulsar wind nebula~\citep{Sarin2022_mag, OmandSarin23}. 

Again, through the use of dependency injection, these models can be easily modified to capture different physics. We also provide an interface to supernova models implemented in \texttt{SNCosmo}~\citep{Barbary2022}, which further amplifies the library of supernovae models available in \redback{}. In future releases, we will be adding surrogate models to hydrodynamical/radiative transfer simulations of interaction powered supernovae, among other models.
\paragraph*{Engine driven transients:}
Distinct from the magnetar-driven supernovae models described above, we also provide a general class of magnetar driven models. Such models aim to capture the emission that would be produced in a magnetar-driven kilonova or a magnetar-driven fast blue optical transient~\citep{Drout2014, Arcavi2016}. Several different models are implemented such as those that capture the dynamical evolution of the nascent neutron star~\citep{Sarin2022_mag} or the dynamical evolution of the ejecta~\citep{Metzger2014, Sarin2022_mag}. 
We also include models with relativistic considerations~\citep{Yu2013, Sarin2022_mag}, non-vacuum dipole spin~\citep{Lasky2017}, and models with variation in their treatment of the thermalisation efficiency or gamma-ray leakage~\citep{Wang2015, Sarin2022_mag}. We also include an implementation of the trapped magnetar model that has been suggested as an explanation for the enigmatic fast X-ray transient, CDF-S XT1~\citep{Sun2019}. 
In the future, we will add models to capture energy injection from fallback accretion onto a central black hole.
\paragraph*{Millisecond magnetar:}
Ever since the launch of \textit{Neil Gehrels Swift Observatory}~\citep{Gehrels2004}, the origin of the X-ray afterglows of gamma-ray bursts has been a long source of debate. In particular, features referred to as the internal and external plateaus are difficult (although not impossible) to explain within the standard picture of synchrotron emission from a jet interacting with the ambient medium. These plateaus are readily explained as the bare or processed spin down from a highly magnetic, rapidly rotating newly born neutron star i.e., a millisecond magnetar.

In \redback{}, we provide several implementations of millisecond magnetar models, such as early models which assumed the neutron star only spun down through vacuum dipole radiation~\citep{Zhang2001, Rowlinson2013}, to extensions that included a variable braking index~\citep{Lasky2017}. We also provide models which include a collapse time~\citep{Sarin2020_collapsing}, to capture lightcurves when the neutron star undergoes a delayed collapse to a black hole. 
The above models all implicitly assume that the observed emission is a constant factor of the real spin-down power of the neutron star. In reality, it is difficult to assume that this factor will be constant in time and be the same for different environments/ejecta properties. To capture this behaviour, some other models have been developed which account for this changing efficiency by accounting for the radiative losses at the interface between the jet and interstellar medium~\citep{Dall'Osso2011, Sarin2020_radiative}, these models are also implemented in \redback{}. Similar to the extension in physics of how emission is generated, the assumption that a neutron star spins down with a constant braking index is also simplistic, we therefore include models where the braking index is a time-dependent value conditioned on the evolution of the angle between the spin and magnetic field axes~\citeg{SasmazMus2019, Sarin2022_mag}. 
\paragraph*{Tidal disruption events:}
Tidal disruption events occur when a star in a galactic nucleus approaches a supermassive black hole (SMBH) and is sufficiently close to be torn apart by tidal forces~\citep{Hills1975}. Many models for tidal disruption events exist which have different assumptions of how the optical/UV lightcurve is produced. For example, some models assume that the optical/UV lightcurve directly tracks the fallback rate~\citep{Guillochon2013, Guillochon2017, Mockler+19}, consistent with the light curve decay slope of $L \propto t^{-5/3}$ expected for complete disruptions~\citeg{Guillochon2013}. Other models assume that the disrupted material does not circularise rapidly and instead the light curve is powered by stream-stream collisions~\citep{Piran2015, Ryu+20b, Ryu+23}. Recent numerical simulations have shown that disrupted material does indeed circularise rapidly~\citep{Steinberg&Stone22} but this need not lead to rapid feeding of the SMBH, instead the material forms a quasi-spherical pressure supported envelope rather than in an accretion disk~\citep{Metzger2022}. 

Motivated by these different assumptions, in \redback{}, we provide two primary sets of models; the cooling envelope model described in~\citet{Metzger2022} and \citet{SarinMetzger23}, which models the optical/UV emission from a cooling envelope and a more fallback rate inspired model similar to \mosfit{}~\citep{Guillochon2018, Mockler+19}. 
In future versions, we will add models that describe the lightcurve from stream-stream collisions and surrogates that directly emulate the lightcurve produced by radiative transfer simulations.
\paragraph*{Shock-powered models:}
The emission produced via shocks is diverse and an important ingredient for many different transients, such as the early cooling that may occur in a supernova or kilonova ejecta~\citep{Piro2018}, the shock powered emission when a blastwave interacts with the preceding material such as supernova explosions with circumstellar interaction~\citep{Margalit2022_model, Margalit2022_csm}. Or the synchrotron emission produced in mildly-relativistic blast waves with both thermal and non-thermal electrons~\citep{Margalit2021_thermal}. In \redback{}, we provide an individual model for each of these processes, to be used independently or added onto any other \redback{} model. 
\paragraph*{Prompt gamma-ray burst:}
The mechanism that produces the high-energy gamma-ray emission in gamma-ray bursts is unclear. However, the prompt emission lightcurves of gamma-ray bursts are often analysed to look for signatures of periodicity~\citep{Hubner2022, Chirenti2023}, lensing~\citep{Paynter2021}, or to characterise the observations into different GRB subtypes. In \redback{}, we provide five models for gamma-ray burst lightcurves to facilitate this research.
\subsubsection{General purpose models}
\paragraph*{Generic models:}
While physical intuition is often the highest priority when performing inference, sometimes we require a model that is robust, flexible and will fit all our observations. Such models can often form the basis of more physically-motivated models or just be used to directly gain insight into the population. In \redback{}, we provide several phenomenological models to address this aim, from models which mimic a Gaussian rise, to an exponential rise and power law decay, to broken power laws with one to six components. As these models have no physics, they are often orders of magnitude faster to evaluate and fit than the physical models described above, making them particularly practical as a way to screen transient candidates.  
\paragraph*{\redback{} surrogates:}
All of the models described above rely on an analytical or semi-analytical model prescription for the physics dictating the lightcurve. Although such models are incredibly useful for getting insight into different transient phenomena, they likely make simplified assumptions which may not be suitable to draw accurate inferences into observations. 
In an independent package, \surrogate{}, which has a direct interface to \redback{}, we provide a library of models which are machine learning surrogates to numerical simulations. At present these models are restricted to surrogates of kilonovae simulations~\citep{Lukosiute2022, Kasen2017, Bulla2019}. All models in \surrogate{} seamlessly integrate into \redback{} and can be used like any other model implemented in \redback{}. 
In future releases, we will provide surrogates for hydrodynamical/radiative transfer simulations of many different transients as well as an interface to build your surrogate from a grid of simulations. 

\paragraph*{Joint Afterglow/Kilonova/Supernova:}
Observations of supernovae in afterglows~\citep{Zeh2004, grenier2015, Cano2017} and more recent infrared excesses consistent with a kilonova in some GRBs~\citep{Tanvir2013, Lamb2019, rastinejad22,Levan2023} have motivated jointly fitting the broadband afterglow alongside a kilonova or supernova component. In \redback{}, we provide three such joint models to enable joint fitting. In particular, a tophat afterglow with an Arnett model, to jointly fit a wide variety of GRBs with supernovae, and two models for jointly fitting a kilonova, one using a two-component kilonova following~\citet{Villar2017} and another following the heating-rate model~\citep{Hotokezaka2020} with a simple tophat afterglow. 

We note that to keep a consistent data generation method, these models can only be fit in flux density, requiring the assumption that optical band pass magnitudes are approximately equivalent to the flux density at the band pass effective wavelength. 
We further emphasize that these models are simply adding the prediction of the two emission processes and do not capture the complicated physics e.g., the interaction of the jet with the ejecta that may significantly alter the overall lightcurve~\citep{Klion2021, Nativi2021}. We also note that while the above options are limited in variety, the choice is motivated by both the simplicity (less parameters to fit) and flexibility of the models. Users of \redback{} can replace each of the individual components with a different model implemented in \redback{} or their own model. We provide an additional, simple joint model interface that enables users to use any other \redback{} afterglow or kilonova/supernova model, only requiring the user to pass a string referring to the model they wish to use. 
\paragraph*{Gaussian process base model:}
While the large diversity of models in \redback{} offers a lot of opportunity that one model might explain observations sufficiently well. Transient phenomena is quite often too complicated, and often the data we observe has underlying processes e.g., periodicity, correlated noise or unmodelled physics that can not be captured analytically or not understood \textit{a priori}. To provide even more flexibility and as a better estimate of uncertainty and fitting procedure in the presence of correlated noise, we provide a generic interface to Gaussian processes in \redback{}. In particular, every model in \redback{} can be used as a mean model for Gaussian process kernels implemented in \texttt{George}~\citep{Foreman-Mackey2015} and \texttt{celerite}~\citep{Foreman-Mackey2017}.
\paragraph*{Phase and Attenuation models:}
All \redback{} models are written with the assumption of no attenuation and that the transient time observations are since the transient started (i.e., that the time of the explosion is known). In practice, these assumptions are mostly incorrect. Therefore, we provide an interface which for all \redback{} models can make the time in reference to an unknown start time (which can be added as a parameter to sample) and/or add attenuation which can be added as a parameter to be estimated by sampling. The attenuation is handled through the {\tt extinction} package~\citep{Barbary2016}. We note that \redback{} assumes all photometry has already been corrected for Milky Way extinction before creating a \transient{} object. However, if not, the user can do this through the {\tt extinction} package alongside online resources to gather the Milky Way extinction along the line of sight of the transient. 
\subsubsection{Acknowledgement of models}
Many of the models implemented in \redback{} are implementations of models that have been described previously in the literature or exist as an interface to another open-source package. To ensure these previous works are adequately acknowledged and facilitate development we provide a simple one line attribute to all models that will provide a reference to the NASA ADS page for the paper describing the model or the software that originally implemented this model.

\subsection{Simulation}\label{sec:simulation}
A key requirement for inference workflows is the ability to test pipelines on realistic synthetic data. To wit, we have created a simulation module in \redback{} to create lightcurves for transients that can be loaded in a transient object and used in inference. Specifically, we provide three classes. 

\begin{enumerate}
    \item \textbf{A generic simulation interface} that can be used to create simulated data for any type of transient. In this module, the time, observed filters/frequencies are sampled randomly from user inputs and added to a user-specified noise level. This generic interface can be used for any \redback{} model and is appropriate for generating Target of Opportunity (ToO) style of observations rapidly. 
    \item \textbf{A more detailed simulation interface} specifically for optical transients to be used for producing lightcurves from real or user-generated surveys/telescopes. Specifically, here we use official table of pointings for ZTF and the Vera Rubin Observatory (provided in \redback{}), which describe the pointings of the telescope, the limiting magnitude, cadence of filters and other properties. Users can also build a pointings table with minimal inputs and design their own survey or provide a table of pointings from an survey not implemented in \redback{}. This allows any \redback{} or user-provided model to be used to generate realistic survey lightcurves and not only validate their inference methodologies but also understand constraints from survey lightcurves or optimize survey design.
    \item \textbf{A full survey } Here a user provides a rate, a survey duration and a \redback{} model and prior (described in detail below) and a full survey is generated with events drawn according to the rate, placed isotropically in the sky and uniformly in co-moving volume. The detected/not-detected events are tracked and this can be used to understand the detectable fraction of events and how that is affected by the population properties of the transient and survey strategy. 
\end{enumerate}

We note that we assume a circular field of view for simulating real surveys in \redback{}. 
This is incorrect for surveys such as ZTF, which has a rectangular field of view and a circular field of view could underestimate the rate of transient detections if adopting a circular field of view. 
However, this approximation is likely not a concern, in ZTF, the fields are fixed to the same sky coordinates with no dithering, which provides uniformity and more accessible reduction and background subtraction. Nevertheless, the transients landing on the gaps between the CCD quadrants are consistently lost. This results in an $\approx 15\%$ loss of the effective area. 
In \redback{}, we approximate the 47 sq.deg rectangular field of view of ZTF as a perfect inner circle of 36 sq.deg, corresponding to a loss of $\approx 20\%$, which is a reasonable approximation for most studies given significant uncertainties on rates and source properties. For, LSST, this is not a concern as the Rubin field of view can be well-approximated as a circle. We will improve the treatment of different surveys' focal plane geometry in future releases. This \simulation{} interface can also be used to optimize survey strategies and design for different transients or specific science goals. 

\subsection{Inference}
The key aim of \redback{} is to enable Bayesian inference on electromagnetic transients. For inference, \redback{} leverages the interface to \bilby{}, which provides a wrapper to many open source sampling software. With this interface a user of \redback{}, simply needs to 1) specify an implemented sampler as a string ($16$ samplers are implemented at time of writing), 2) write a prior (or use the default for the model), 3) specify a likelihood (chosen by default unless specified) and then 4) fit a model. 
In this paper, we assume familiarity with Bayesian inference but we refer readers who are beginning in this field to~\citep{Mackay2003, Hogg2010, bilby1} and references therein. 

\subsubsection{Likelihoods}
Likelihoods in \redback{} are chosen by default and apart from the exception of photon count data (which uses a Poisson likelihood), are by default, Gaussian. However, the modular interface means that users can change the likelihood used with one line of code to another \redback{}-implemented likelihood (there are several to choose from) or write their own and use that instead. This flexibility enables \redback{} to be useful to both advanced users who wish to model the likelihood more accurately and users who simply wish to fit a transient.

\subsubsection{Priors}
To obtain a posterior in Bayesian inference, we require a prior. For all \redback{} implemented models we provide a default prior, this prior is typically broad and uninformative. \redback{} priors are written in the same way as \bilby{} priors and are effectively a dictionary with keys corresponding to each prior. Many prior distributions are implemented but users can also implement their own which they either write mathematically or provide a grid of the prior that can be used to build an interpolant. \redback{} also provides access to conditional priors to write priors on parameters that depend on one another. Many astrophysical models also have constraints, for example in engine-driven models we always want to ensure that the energy in the ejecta does not exceed the energy budget of the engine or that our flux does not exceed a known upper-limit/non-detection. These conditions can be placed on any prior as a \texttt{Constraint}, which will ensure that any prior draw does not violate any constraints. All \redback{} priors can also be sampled from with one line of code to enable users to better understand the prior distributions.

\subsubsection{Samplers}
There are many advantages to being able to choose from a list of samplers (with no additional overhead beyond changing one line of code), for example, several samplers come with the ability to do parallel processing, which can dramatically improve run times. Some samplers also have the ability to resume from checkpoints and produce regular diagnostic plots that can be used to verify progress. There are also large differences in the algorithm of certain samplers, beyond the general distinction between nested sampling and Markov Chain Monte-Carlo, with some algorithms better suited to one type of transient than another. 

For \redback{} specifically, we use the \dynesty{}~\citep{Speagle2020} by default, but we regularly find that \pymultinest{}~\citep{pymultinest} and {\sc{nestle}}\footnote{\href{ http://kylebarbary.com/nestle/}{http://kylebarbary.com/nestle/}} give similar posteriors for significant shorter run times. However, the latter tend to be less robust at dealing with a complicated parameter space. 
A full sampler comparison is beyond the scope of this paper but we strongly encourage users to perform inference with multiple different samplers, both to gain a better understanding of the parameter space, what algorithms perform best and as a cross sampler validation to ensure that their results have converged.

\subsection{Format of results}
After a fit, \redback{} returns a homogeneous \result{} object. This object is the same for any type of transient analysed. The object is also saved locally (in a machine-readable \texttt{json} file by default) either with a user-specified location/label or as a sub folder with the name of the model in a folder that is the name of the type of transient analysed (by default). 
The \result{} object contains several attributes needed for diagnosis, such as a pandas data frame of the posterior values, alongside metrics (depending on the sampler) such as the Occam factor, the Bayesian evidence, the number of likelihood evaluations, the priors used in the analysis and additional metadata which includes a copy of the \transient{} object used in the fit. The \result{} object also contains several methods, from convenience functions to obtain the credible intervals and latex strings for the constraints on all parameters, to plotting the corner or lightcurve and multi-band lightcurve plots with the data and the fit. 
The \result{} file can also be shared and loaded in \redback{} to enable users to share their analysis or work across multiple machines. We note that the \redback{} \result{} object inherits from the \bilby{} \result{} object, inheriting additional useful methods and diagnostics such as the ability to importance sample or make a percentile-percentile (PP) plot~\citep{ppplot} to validate an inference workflow. 
\subsection{Plotting}
In \redback{}, all plotting methods are implemented in a specific \texttt{plotting} module. However, we note that the access to these methods is through the \transient{} and \result{} objects. In particular, we provide interfaces to plot the observations themselves, the fit to single or multi-band photometry as random models drawn from the posterior or as a credible interval and a residual plot. The different \redback{} plotting functionality is demonstrated in Sec.~\ref{sec:examples}.
To simplify modification of \redback{} plots, all plotting methods return the \texttt{matplotlib} axes, which can allow users to change things such as the axes labels/fontsize/scale/limits or plot something extra on the same plot. Furthermore, users can also pass their own \texttt{matplotlib} figure and axes to \redback{}, enabling multi-panel lightcurve plots or a customized size. The \texttt{plotting} module also uses dependency injection and keyword arguments for several settings which can be used to change many features of the different plots. Users can also replace the \texttt{plotting} module to be more specific to their needs or call the model themselves to plot what they would like. 

\begin{figure*}
    \centering
    \includegraphics[width=1.00\textwidth]{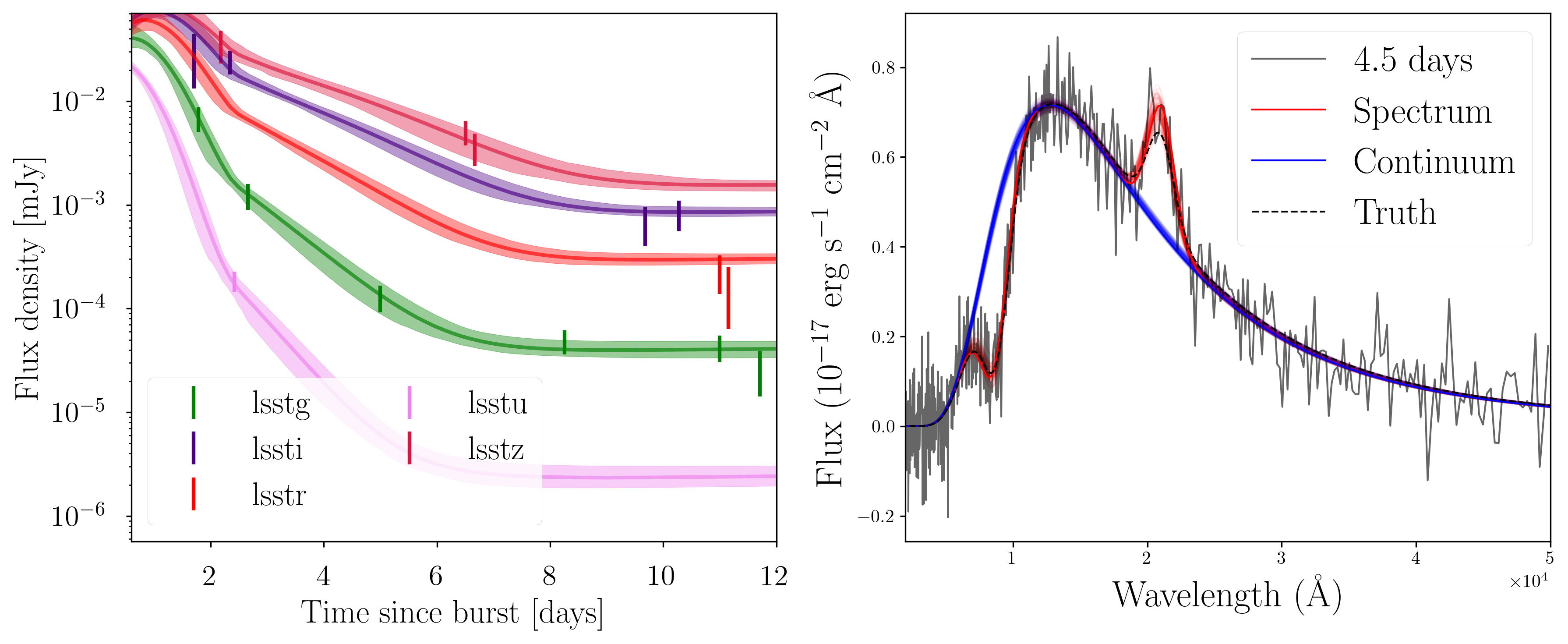}
    \caption{Simulated photometric data (left panel) with colours corresponding to different LSST filters and spectroscopic data (right panel) in black for a kilonova. In the left panel the shaded band shows the $95\%$ credible interval fit to the photometry. On the right panel, we show the predicted continuum flux (in blue) and total spectrum (red) from $100$ randomly drawn points from the posterior, alongside the true input in black dashed lines.}
    \label{fig:spectrum_photometry}
\end{figure*}

\subsection{Analysis}
Separate from the main modules provided in \redback{}, we include an \texttt{analysis} module that can be used to set up the different workflows or make additional diagnostic plots for some models or calculate prior/posterior predictions for other properties.
For example, here we provide a method to plot lightcurves generated by a user-provided set of parameters on top of the `plot\_multiband' or `plot\_data' generated plots to get a sense of the appropriate prior for fitting or build intuition about a model. Alongside this, we provide methods to plot the spectrum generated by \redback{} model, or additional posterior predictive plots such as of the evolution of the nascent neutron star. 
In the future, we will add more diagnostic analysis methods and encourage \redback{} users to contribute with typical diagnostic plots and calculations of their favourite transient. 
\subsection{Directory Structure}
By default, the \redback{} directory structure is set by the type of transient, the name of the transient and the model used in fitting. For example, if one downloads the data for the kilonova, AT2017gfo, this data will be saved to a folder called \texttt{kilonova} in the current working directory. If a user then loads this data and fits with a model called \texttt{redback}, then the \result{} file alongside all plots and sampler-specific diagnostics will be saved to a folder within \texttt{kilonova} with the model name. 
This behaviour can be changed in two primary ways. 1) The user can specify an \texttt{outdir} and \texttt{label} when running the fit (see below) which will save the result to folder \texttt{outdir} with the \texttt{label} prepended to any output. 
2) The user can change the \texttt{name} attribute of the \transient{} object. Which will change the label that is prepended to any output file but keep the default directory structure. We note that any \result{} files generated by a non-default directory structure can simply be loaded up by specifying the path, while plotting locations can also be specified via the typical method of \texttt{matplotlib}.

\section{Joint analysis of spectrum and photometry}\label{sec:joint_spectrum_photometry}
With the software's design objectives and overview out of the way. We now turn towards a new application enabled by \redback{}. As we described in the introduction, it is becoming increasingly common for electromagnetic transients to have extensive spectroscopic and photometric observations. However, photometric analyses and spectroscopic analyses are often performed independently. Typically, the spectrum is often used primarily for the identification of a redshift and to identify the type of transient and later potentially specific emission lines. Meanwhile, the photometry is left for estimating the properties of the transient, such as the ejecta masses in supernovae and kilonovae or the black hole mass in tidal disruption events.

It is understandable that currently, analysis of the spectrum and photometry is performed separately, given the high computational cost of detailed spectral models and analytical/semi-analytical models that work on photometry but fail to capture the details of a spectrum. 
However, it is often the case that separate analyses of photometry and spectrum can provide contradictory information. For example, some supernovae observations where the photometry are often better described purely by $^{56}$Ni decay while the spectrum has tell-tale signatures of interaction with circumstellar material. Each independently suggests different quantities of ejecta, making it difficult to understand the properties of supernovae explosions and can sometimes even change the interpretation of specific events~\citeg{Schulze2023}. Or the case of the kilonova, AT2017gfo, where the spectrum at $1.4$ and $4.4$ days is best described by electron fractions~\citep{Gillanders2022} inconsistent with those used to fit the photometry~\citep{Villar2017}. Such contradictions are likely down to modelling limitations. However, it is critical we understand which of the estimated properties are more robust, where our modelling could be improved and what the photometric and spectral observations are jointly telling us. Joint analysis can also provide significantly more powerful constraints by breaking degeneracies present in the independent analyses and thereby improving our estimation of the transient properties. This has important consequences as, ultimately, we aim to use the estimated parameters of the explosion to answer fundamental questions in physics and astrophysics.

We now describe how \redback{} can be used to jointly fit the spectrum and photometry of a kilonova. For the purposes of this demonstration, we choose a simplified simulated spectrum and photometry to ensure we can validate the entire process. This is a specific example of workflow B, described in Sec.~\ref{sec:design}. We simulate target-of-opportunity observations of a hypothetical kilonova, AT2025ixp, observed by the Vera Rubin observatory through the \redback{} \simulation{} module. In particular, we use the two-component kilonova model implemented in the \transientmodels{} subpackage following~\citet{Villar2017}. We then evaluate the spectrum at $\unit[4.5]{d}$ from this model, by calling the model with an additional keyword argument to change the output format of the model. Assuming this model only captures continuum emission, we add an additional absorption and emission line at $8800$ and $21000$ \AA{} respectively. Here, we model both spectral lines as a Gaussian, mimicking their Doppler broadening due to the high-velocity kilonova ejecta. We add Gaussian noise to the total spectrum (spectral lines and continuum emission) comparable to noise in the X-shooter spectrum of AT2017gfo~\citep{Pian2017, Smartt2017}. With the data generated we create an instance of the kilonova \transient{} object. We then independently fit the spectrum and photometry and jointly fit both together using the \pymultinest{} sampler~\citep{pymultinest} through the \redback{} interface, specifying a Gaussian likelihood via the \likelihood{} module and broad uninformative priors via the \prior{} module.

In Fig.~\ref{fig:spectrum_photometry} we show the results from our analysis. In particular, in the left panel, we show the data in multiple LSST filters alongside the $95\%$ credible interval from our fit to the photometry. In the right panel, we show the simulated spectrum at $\unit[4.5]{d}$ (in black) alongside our fit, showing the continuum emission in blue and the full spectrum, including absorption and emission lines in red. In both cases, we see we can fit the observations well, correctly recovering the input. 

\begin{figure}
    \centering
    \includegraphics[width=0.95\columnwidth]{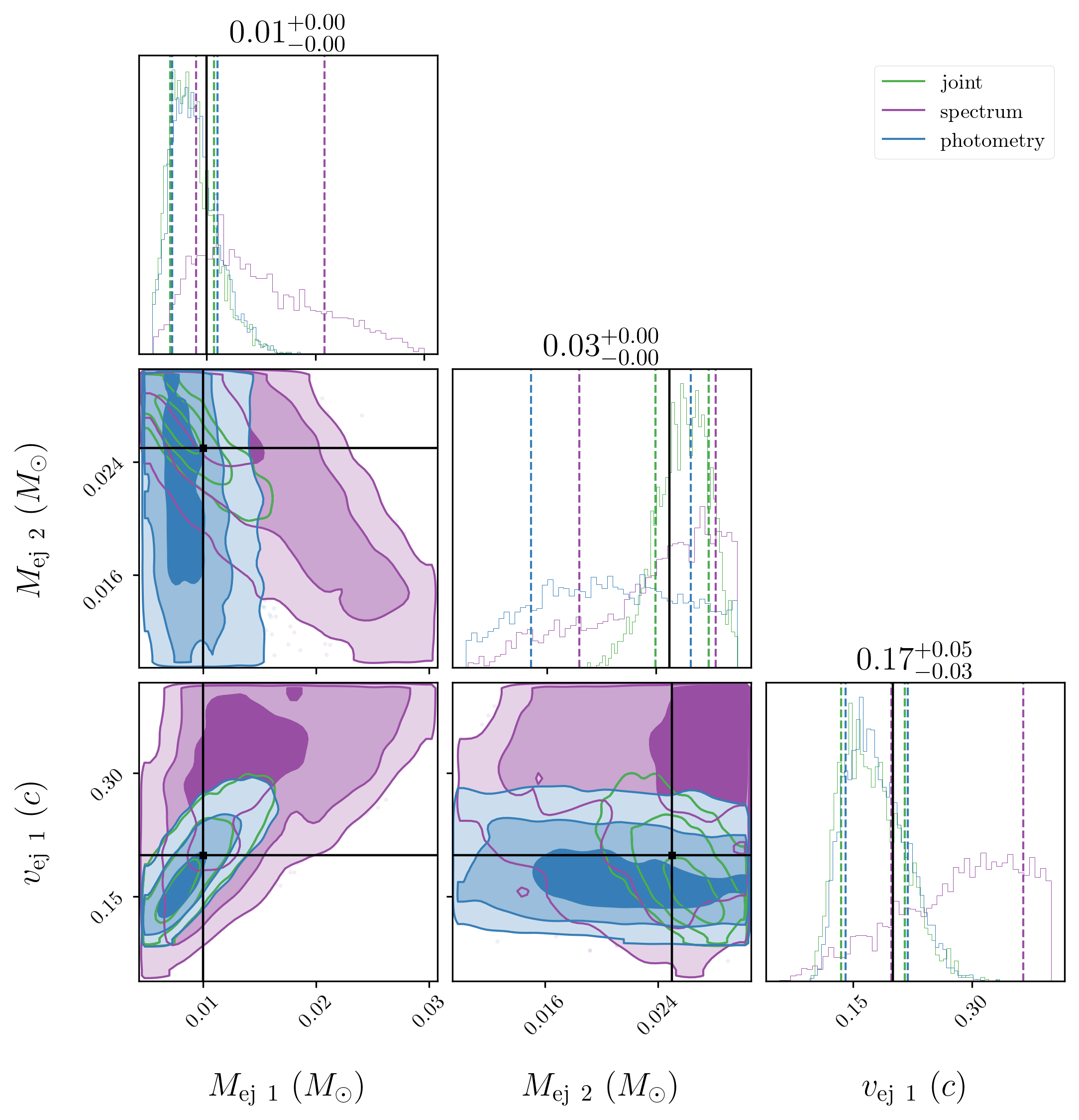}
    \caption{Posterior distribution on the component ejecta masses, M$_{\rm ej,1}$ and M$_{\rm ej, 2}$}, and bulk ejecta velocity of the first component, v$_{\rm ej,1}$ for the joint fit and the photometry and spectrum independently.
    \label{fig:joint_specphot_corner}
\end{figure}

In Fig.~\ref{fig:joint_specphot_corner}, we show the posterior distributions on multiple parameters of the two-component kilonova model from the independent spectrum and photometry fits and the joint fit. These posteriors highlight the power of joint analysis, while all analyses recover the true input (indicated by black lines), the joint spectrum and photometric fit do so with significantly more precision by breaking the degeneracy in the independent photometric and spectroscopic analyses. For example, the precision of the second ejecta component's mass and velocity improves from a precision of $29$ and $15\%$, respectively, from the independent fit to the photometry to a precision of $8$ and $4\%$. This boost to precision has several important consequences as kilonova properties have been previously shown to be useful for constraints on the behaviour of nuclear matter~\citep{Pang2022}, constraints on the Hubble constant~\citep{Perez-Garcia2022}, while offering better precision to ultimately understand how these explosions work. 

The above example is a demonstration of one of the unique capabilities of \redback{}: a cohesive, single framework analysis of spectrum and photometry that is facilitated by the modular design of \redback{}. Specifically, \redback{} enables stitching together different functional modules for different problems. In particular, we can jointly, and independently fit both the spectrum and photometry by setting up the relevant functional modules in distinct ways. Similarly, we can simulate the two types of data in question, enabled by the design implementation of all models such that they can be evaluated for arbitrary inputs, times and return outputs in multiple formats, alongside the implementation of the \simulation{} and \transient{} functional modules, where the former can be used to generate synthetic observations for distinct data types, while the latter has the required flexibility to handle such distinct data.

\section{Future development}\label{sec:future}
As we continue to drive progress in transient astronomy, we develop newer and better models for transients and make improvements to how we treat the data. This paper marks version 1.0 release but \redback{} will be further developed to keep pace with the developments in modelling and treatment of data. 

One of the primary aspects that will be improved are the models implemented in \redback{}. In particular, we are currently implementing models for interacting supernovae and fast-blue optical transients, from semi-analytical models of shocks produced by interacting shells~\citep{Margalit2022_model}, to surrogates of radiative transfer simulations~\citep{Khatami2023}. We are also improving some of our models of afterglows for better treatment of reverse shocks and to make them more computationally efficient. 
We will soon implement model for $r$-process nucleosynthesis from collapsars~\citep{Barnes2022, Anand2023}. 
On longer timescales, we will implement models with better spectral modelling, enabling joint fitting of the spectrum and photometry. 

Alongside improvements and addition of models, we will further develop \redback{} for more practical purposes, for example, providing a generic interface in \surrogate{} to allow users to make their own surrogate from a grid of simulations and newer likelihoods that better describe the data generation process. We will also be further developing the \simulation{} module to improve our treatment of focal plane geometry. On longer timescales, we will add some GPU implementations of models to enable rapid inference and an API to download and process data from the Fermi catalog~\citeg{vonKienlin2020}. 
\section{Conclusion}\label{sec:conclusion}
Realising the rich promise of the large transient data expected from new observing facilities such as the Vera Rubin Observatory and ULTRASAT~\citep{Shvartzvald2023} requires us to confront such data with models describing the different transient phenomena. 
This requires fast, reliable, open-source code that is both accessible to newcomers to the field and modular such that it can be adapted to be the powerhouse required by experts. Here, we have described \redback{}, a Bayesian inference software package for end-to-end for parameter estimation and interpretation of electromagnetic transients. 

\redback{} is an engine for simulating realistic transients and inferring their properties enabling end-to-end analysis and validation of inference workflows. Furthermore, one can also use this software to understand how to optimize survey strategies/design or understand the selection function of different telescopes/surveys. 
\redback{} is also fully Bayesian, enabling the vast advantages of this statistical paradigm such as model selection, importance sampling, and Bayesian hierarchical modelling. 
We reemphasize here that \redback{} is object-orientated, enabling users to input their own model, priors, and data without needing to edit the source code, and simply replace any functional module of \redback{} with their own code. The interface to \bilby{} also provides access to a large variety of samplers enabling validation across samplers and a simplistic interface for multi-messenger analysis for joint events such as GW170817~\citeg{Radice2018,Coughlin2019, Gianfagna2023}. 
These design objectives address many of the limitations of previous open-source packages for electromagnetic transients.

In this paper, we have described the overall design of \redback{}, a new scientific application where we jointly fit the spectrum and photometry of a kilonova. This holistic look at a complete transient dataset offers the opportunity to both increase the precision of our constraints and confront contradictions that may emerge when interpreting only one type of data. For the specific case of a kilonova, we show how joint fitting can dramatically improve the precision of the inferred ejecta masses, increasing the value of each event for constraints on the equation of state. Or also remove biases inadvertently caused by fixed opacities in photometric analyses that are inconsistent with the spectrum. In the Appendix, we provide additional examples demonstrating the functionality and usability of the software in various applications and a general interface. 

As discussed in Sec.~\ref{sec:future}, we will continue to further develop \redback{}, including the addition of newer models and additional functionality. 
\redback{} has already been used in previous publications such as inference on tidal disruption events~\citep{SarinMetzger23}, analysis of SN~2018ibb~\citep{Schulze2023}, magnetar-driven kilonovae and supernovae~\citep{Sarin2022_mag, OmandSarin23}, GRB afterglows~\citep{Sarin2021_cdf, Sarin2022_blt}, and to infer joint GRB and kilonovae observations~\citep{Levan2023}, demonstrating the flexibility of the software. A more comprehensive comparison of results for different transient catalogs is underway alongside interpretation for other transients. 
\section{Acknowledgments}
We thank Duncan Galloway, Phil Macias, Dougal Dobie, Karelle Siellez, Fiona Panther, Jacob Wise, Ariel Goobar, and Miti Patel for helpful comments on the software or manuscript. 
We are grateful to several members of the astronomical community for releasing their transient models as modular open-source software, many of these models form part of the available models in \redback{}. 
We hope the practice of releasing open-source software gathers more pace in transient astronomy. We also acknowledge the work done by numerous people to maintain public catalogs such as \swift{} Data Centre, OAC, Fink, and Lasair. 

N. Sarin is supported by a Nordita Fellowship. Nordita is supported in part by NordForsk. 
S. Schulze and A. Sagu{\'e}s-Carracedo acknowledge support from the G.R.E.A.T. research environment, funded by {\em Vetenskapsr\aa det},  the Swedish Research Council, project number 2016-06012.

A part of this work used computing facilities provided by the OzSTAR national facility at Swinburne University of Technology. 
The OzSTAR program receives funding in part from the Astronomy National Collaborative Research Infrastructure Strategy (NCRIS) allocation provided by the Australian Government. 

The \redback{} package makes use of the standard scientific \python{} stack~\citep{scipy, numpy, pandas}, {\tt matplotlib}~\citep{matplotlib}, and {\sc corner}~\citep{corner}, for the generation of figures, and {\sc astropy}~\citep{astropy1,astropy2, AstropyCollaboration2022} for common astrophysics-specific operations. 
\redback{} makes use of \bilby{}~\citep{bilby1, bilby2} to provide an interface to different sampling algorithms and for evaluating prior distributions. \redback{} uses {\sc SNCosmo}~\citep{Barbary2022} for filter definitions and calculations of magnitude from SEDs, {\sc extinction}~\citep{Barbary2016} for extinction corrections. And {\sc requests} and {\sc selenium} for downloading data from catalogs. 

\section{Data Availability}
The software package along with example scripts for all analysis demonstrated in this manuscript alongside a plotting notebook to generate all the plots as well as other examples are available at \redbackurl{}. The specific \result{} objects for each of the analyses presented here are available at \url{https://doi.org/10.5281/zenodo.8273145}. \redback{} is available on {\sc{pypi}}. This paper uses v1.0 release of \redback{} with documentation at \url{https://redback.readthedocs.io/en/latest/}. The data for all transients is available at the Open Access Catalog~\citep{Guillochon2017} gathered through the \redback{} \getdata{} module or hosted at \redbackurl{}. 


\bibliographystyle{mnras}
\bibliography{ref}

\appendix
\onecolumn 
\section{General interface}\label{sec:interface}
We now describe the general interface for \redback{}, for example how to download and load data, simulating a transient or calling a \redback{} model with a constrained prior. We note that these sections are not exhaustive demonstrations of the \redback{} API and merely show some demonstrative functionality. Full API documentation is provided at \redbackdocs{}.

\subsection{Getting data}\label{sec:gettingdata}
As mentioned in Sec.~\ref{sec:overview}, \redback{} provides an API to download and process data from multiple catalogs. This data is saved as a human readable file and returned as a pandas data frame. In particular, 

\begin{lstlisting}
import redback 

# FINK 
name = "ZTF22abdjqlm"
data = redback.get_data.get_fink_data(transient=name, transient_type="supernova")

# LASAIR
transient = "ZTF20aamdsjv"
data = redback.get_data.get_lasair_data(transient=transient, transient_type="supernova")

# Open Access Catalog
tde = "PS18kh"
data = redback.get_data.get_tidal_disruption_event_data_from_open_transient_catalog_data(tde)

# BATSE
name = "910505"
data = redback.get_data.get_prompt_data_from_batse(grb=name)

# SWIFT
GRB = "070809"
data = redback.get_data.get_bat_xrt_afterglow_data_from_swift(grb=GRB, data_mode="flux")
\end{lstlisting}

In all function calls we specify the name of the transient we want to obtain the data for and use the relevant class method of the \getdata{} module. For some of these methods we can also specify the type of transient or the type of data to ensure we get the data we want and that it is saved in the appropriate location. We note that \redback{} only processes the AB magnitude data for sources hosting multi-band photometry. This is not a concern for Fink and LASAIR but may result in a loss compared to the open access catalog. However, the raw data file is also downloaded and users can reprocess the data as they wish.
\subsection{Creating transient objects}\label{sec:creatingtransients}
Once we have the data of a transient, there are many different ways to create a \transient{} object. For example, we provide simple class methods to load data that is downloaded from the OAC, FINK, and LASAIR.  

\begin{lstlisting}
supernova = redback.supernova.Supernova.from_open_access_catalogue(name="ZTF22abdjqlm", data_mode="flux")

sn = redback.transient.Supernova.from_lasair_data(name="ZTF20aamdsjv", use_phase_model=True, 
data_mode="flux_density", active_bands=np.array(["ztfr"]))
\end{lstlisting}
Here, the first line creates a \texttt{supernova} \transient{} object from data that was downloaded from FINK. We note that as FINK and OAC have the same data structure, the OAC method can be used for FINK data. Here we have also specified the \texttt{data\_mode} to be \texttt{flux}, which will create the transient object with the \texttt{flux} data mode. Similarly, the second line creates a supernova object but from LASAIR data. However unlike the FINK example, here we specify an active band, which sets all bands apart from the \texttt{ztfr} band to be inactive (not used in fitting), set the \texttt{data\_mode} to be \texttt{flux\_density} and set \texttt{use\_phase\_model=True}. 
The latter condition ensures that the time values we initialise are in MJD, to fit this data we therefore must also sample in the start time of the event. 

We also provide simplified class methods for loading data from \swift{}, BATSE, and the \simulation{} module. In particular, 

\begin{lstlisting}
kn_object = redback.transient.Kilonova.from_simulated_optical_data(name="my_kilonova", data_mode="magnitude")
\end{lstlisting}
Here, we have loaded the magnitude data for a kilonova;  \texttt{my\_kilonova} we generated using the \simulation{} module. \redback{} \transient{} objects can also be constructed directly, for example by loading in a data file and specifying the specific attributes directly. For example,

\begin{lstlisting}
import pandas as pd 

data = pd.read_csv("example_data/grb_afterglow.csv")
time_d = data["time"].values
flux_density = data["flux"].values
frequency = data["frequency"].values
flux_density_err = data["flux_err"].values
name = "170817A"

afterglow = redback.transient.Afterglow(name=name, data_mode="flux_density", time=time_d,
    flux_density=flux_density, flux_density_err=flux_density_err, frequency=frequency)
\end{lstlisting}
This direct construction of a \transient{} object can be done for any other combination of attributes, enabling users to construct a \transient{} object in many different ways. We emphasize that we provide several other class methods than shown here and refer the reader to \redbackdocs{} for the full documentation.
\subsection{Calling a model}\label{sec:callingmodel}
As alluded to in Sec.~\ref{sec:overview}, all \redback{} models exist as \python{} functions and can be called directly on an arbitrary time array and set of parameters. We also provide a convenient look up dictionary to find the function corresponding to a model as well as convenience functions to obtain the relevant citation for the model (for ease of reference and gather additional information about the model) and return an instance of the default \prior{} for the model. 
\begin{lstlisting}
from redback.model_library import all_models_dict 

model = "one_component_kilonova_model"
priors = redback.priors.get_priors(model=model)
priors["redshift"] = 1e-2    
function = all_models_dict[model]
citation = function.citation 

model_kwargs = dict(frequency=2e14, output_format="flux_density")
time = np.linspace(0.1, 30, 50)
sample = priors.sample()
sample.update(model_kwargs)
fmjy = function(time, **sample)
\end{lstlisting}
Here, the first set of code creates the \redback{} prior object from a string referring to a model implemented in \redback{}, we also set the redshift of the prior to be a fixed value, and use a \redback{} dictionary to conveniently get the function corresponding to the model string. The function also has an attribute `citation' that provides a reference for the model. 
The second set of code sets up some additional keywords required by the model such as the frequency we want to evaluate the model at and an output format. We then call the function on a random sample from the prior and arbitrary time array to obtain the flux density (in mJy) corresponding to the specific prior draw.  This simple workflow can be readily changed to draw many more samples from the prior, add a constraint to the prior and draw from the constrained prior, or add/change keys in `model\_kwargs' to change the physics of the model or the output format.
\subsection{Simulating transient}
\begin{figure}
\centering
\includegraphics[width=0.49\columnwidth]{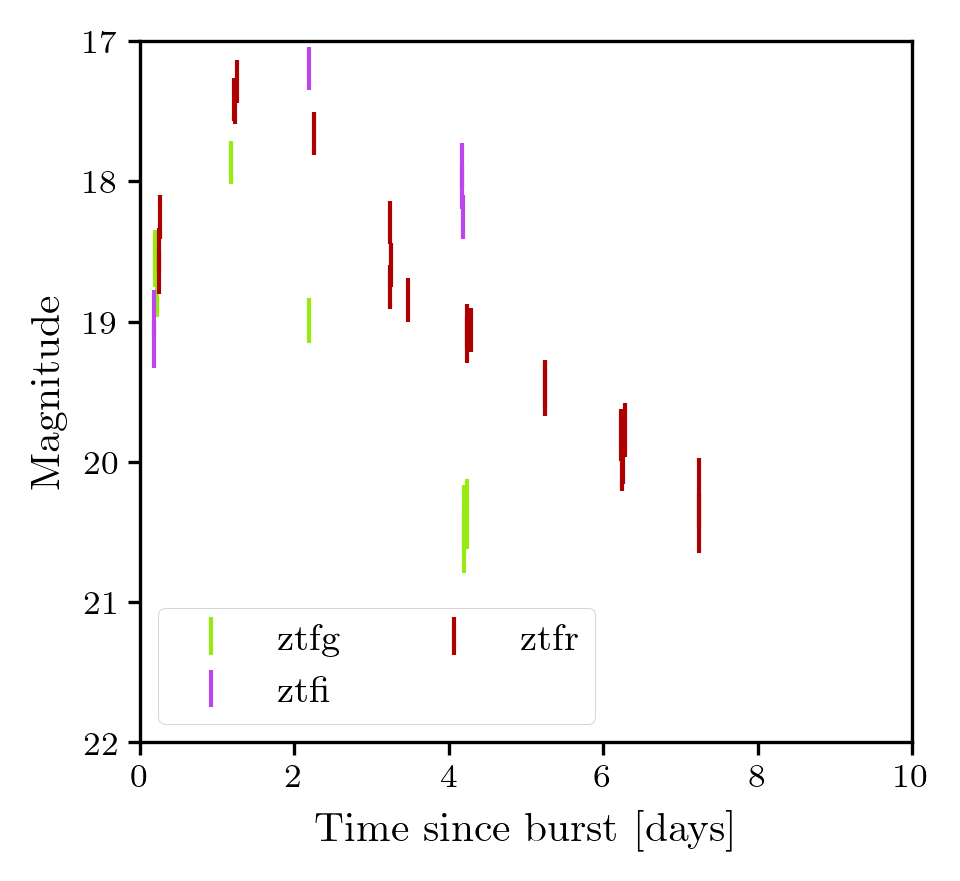} 
\includegraphics[width=0.49\columnwidth]{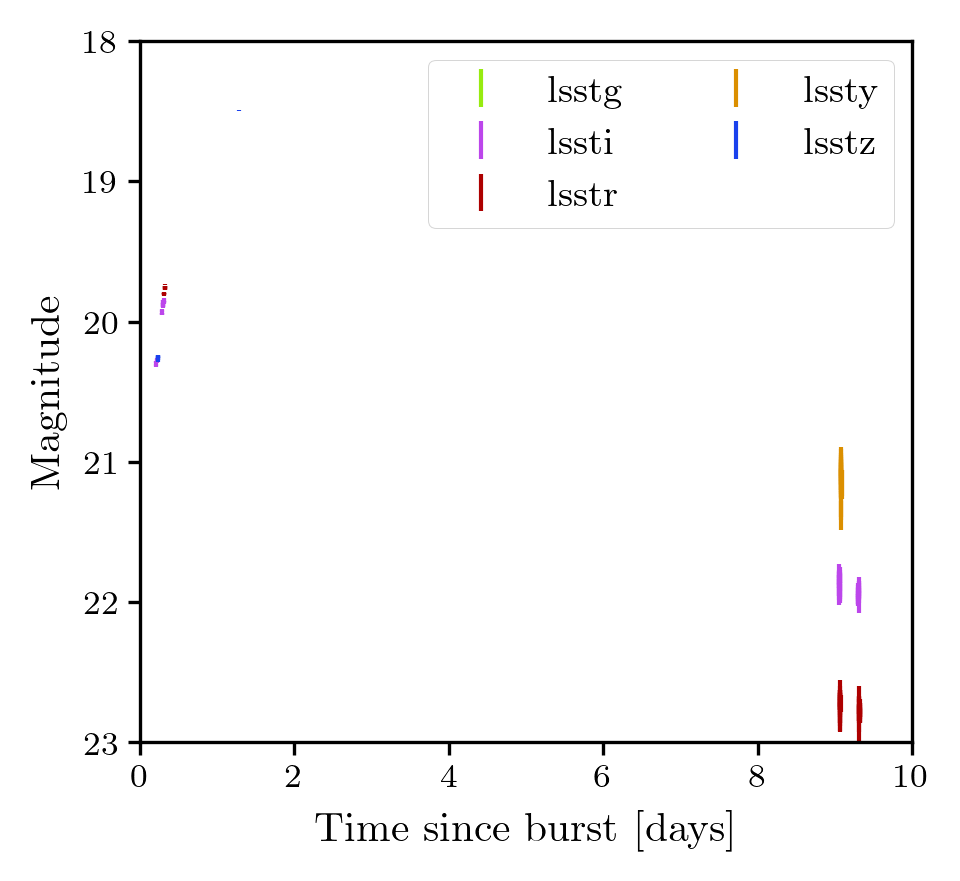}  
\caption{(Left) Simulated kilonova (one component kilonova model) in ZTF and in LSST (Right). We emphasize that aesthetic features such as the colours of the data points, axes limits etc can all be modified by passing in relevant keyword arguments to the \texttt{plotting} methods.}
\label{fig:simulation}
\end{figure}

While the interface described above can be used to simulate data, we also provide a more comprehensive \simulation{} module (described in detail in Sec.~\ref{sec:overview}). For example, generating a simulated lightcurve for a kilonova in ZTF can be done via, 
\begin{lstlisting}
import redback
from redback.simulate_transients import SimulateOpticalTransient

model_kwargs = {}
parameters = redback.priors.get_priors(model="one_component_kilonova_model").sample()
parameters["mej"] = 0.05
parameters["t0_mjd_transient"] = 58288
parameters["redshift"] = 0.005
parameters["t0"] = parameters["t0_mjd_transient"]
parameters["temperature_floor"] = 3000
parameters["kappa"] = 1
parameters["vej"] = 0.2
parameters["ra"] = 3.355395
parameters["dec"] = 0.5820673

kn_sim = SimulateOpticalTransient.simulate_transient_in_ztf(model="one_component_kilonova_model",
    parameters=parameters, model_kwargs=model_kwargs, end_transient_time=15.,
    snr_threshold=5., add_source_noise=True)
\end{lstlisting}
Here, the first set of code specifies the model we want to simulate with and the parameters of the simulated event. Then, we also place it in a part of a sky observable with ZTF (\redback{} will internally randomly place the source within the ZTF observable volume otherwise), then generate a lightcurve with the \simulation{} module. As shown in the Sec.~\ref{sec:creatingtransients}, the simulated data can be easily saved and loaded in a single line of code to create a \transient{} object enabling inference. In Fig.~\ref{fig:simulation}, we show two representative simulated kilonovae in ZTF and the LSST Survey in the Vera Rubin Observatory, demonstrating through a simple example the benefits of the high cadence of surveys such as ZTF for fast transients such as kilonovae. 

We note that this exact interface can also be used to generate survey lightcurves for the Nancy-Grace Roman Observatory or a user-generated survey and for any model implemented in \redback{}, and these examples are available at \redbackurl{}. Furthermore \redback{} also offers the functionality to simulate transients more generically (in a manner more consistent with target of opportunity observations) or simulate a full survey. 
\section{Multi-messenger analysis}\label{sec:multimessenger}
A key advantage of the interface with \bilby{} is to facilitate multi-messenger gravitational-wave and electromagnetic transient analyses. 
Here \redback{} provides the likelihood, model and/or simulated data for the electromagnetic transient and \bilby{} provides the same for the gravitational-wave data. 
Both likelihoods communicate together through the use of a \jointlikelihood{} which combined with a full \prior{}, can be used to perform joint multi-messenger analyses.  

We demonstrate this feature through the observation of a simulated binary neutron star signal, GW231116, observed in O4 alongside a gamma-ray burst afterglow detected in X-rays. 
We note that this workflow can be easily extended to also include an optical/radio afterglow and/or a kilonova. Furthermore, the joint likelihood interface can also be used to jointly fit any two data types, e.g., a spectrum and photometry, both of which could be provided by \redback{} but we leave such examples from this paper for simplicity. This analysis has been performed for GW170817 by multiple groups~\citeg{Gianfagna2023}. 

We start by setting up the data, 
\begin{lstlisting}
import bilby
import redback
from astropy.cosmology import Planck18 as cosmo
from redback.transient_models.afterglow_models import tophat
from bilby.core.prior import Uniform

source_redshift = 0.03
source_distance = cosmo.luminosity_distance(source_redshift).value

gw_injection_parameters = dict(mass_1=1.5, mass_2=1.3, chi_1=0.02, chi_2=0.02, luminosity_distance=source_distance,theta_jn=0.43, psi=2.659, phase=1.3, geocent_time=1126259642.413,
    ra=1.375, dec=-1.2108, lambda_1=400, lambda_2=450, fiducial=1)
\end{lstlisting}
For demonstrative purposes, we assume that the afterglow kinetic energy is some unknown fraction of the total rest mass energy of the binary, alongside the more conventional assumption that the jet is launched along the orbital angular momentum of the binary. These assumptions are not captured by any afterglow model implemented in \redback{}, so we create a new function, wrapping a simple \texttt{tophat} model already implemented in \redback{}.
\begin{lstlisting}
def get_jet_energy(mass_1, mass_2, fudge):
    total_mass = (mass_1 + mass_2)
    return total_mass * fudge * 2e33 * 3e10**2

fudge_factor = 0.04
afterglow_energy = get_jet_energy(gw_injection_parameters["mass_1"], gw_injection_parameters["mass_2"],
fudge=fudge_factor)
grb_injection_parameters = dict(fudge=fudge_factor,
                                theta_jn=gw_injection_parameters["theta_jn"],
                                redshift=source_redshift, loge0=afterglow_energy,
                                thc=0.1, logn0=-1, p=2.2, logepse=-1, logepsb=-2, ksin=1,
                                g0=50, mass_1=gw_injection_parameters["mass_1"],
                                mass_2=gw_injection_parameters["mass_2"])

def grb_afterglow_model(time, redshift, theta_jn, mass_1, mass_2, fudge, thc, logn0, p,
                        logepse, logepsb, ksin, g0, **kwargs):
    energy = get_jet_energy(mass_1, mass_2, fudge=fudge)
    energy = np.log10(energy)
    if "loge0" in kwargs.keys():
        kwargs.pop("loge0")
    return tophat(time=time, redshift=redshift, thv=theta_jn, loge0=energy, thc=thc,
                  logn0=logn0, p=p, logepse=logepse, logepsb=logepsb,
                  ksin=ksin, g0=g0, **kwargs)
\end{lstlisting}
We can now simulate the electromagnetic data using this model following the method outlined in previous sections or by calling the model directly, and then create a \redback{} \transient{} class, alongside an instance of the likelihood. Furthermore, we can set up the gravitational-wave analysis, to reduce the computational cost we use the relative-binning approximation~\citep{Zackay2018, Krishna2023}. We follow the standard \bilby{} relative-binning example for this aspect and do not outline the details here. 
We can also set up the electromagnetic aspect (i.e., the prior and likelihood) via 
\begin{lstlisting}
em_priors = bilby.core.prior.PriorDict()
em_priors["redshift"] = source_redshift
em_priors["thc"] = Uniform(0.01, 0.2, "thc", latex_label = r"$\theta_{\mathrm{core}}$")
em_priors["logn0"] = Uniform(-4, 2, "logn0", latex_label = r"$\log_{10} n_{\mathrm{ism}}$")
em_priors["p"] = Uniform(2,3, "p", latex_label = r"$p$")
em_priors["fudge"] = Uniform(0.01, 0.1, "fudge", latex_label = r"$f_{\mathrm{fudge}}$")
em_priors["logepse"] = grb_injection_parameters["logepse"]
em_priors["logepsb"] = grb_injection_parameters["logepsb"]
em_priors["ksin"] = grb_injection_parameters["ksin"]
em_priors["g0"] = grb_injection_parameters["g0"]

em_likelihood = redback.likelihoods.GaussianLikelihood(x=sim_afterglow.time,
                                                         y=sim_afterglow.flux_density,
                                                         function=grb_afterglow_model,
                                                         sigma=yerr, kwargs=afterglow_kwargs)
\end{lstlisting}
Here, we have first set up a prior on a series of parameters, while fixing some to the injected values to reduce the computational cost of the analysis, and then set up the electromagnetic likelihood, using the \transient{} object attributes.

Once, the electromagnetic and gravitational-wave is set up (i.e., the individual likelihoods and priors), we can simply set up the joint analysis via, 
\begin{lstlisting}
joint_likelihood = bilby.core.likelihood.JointLikelihood(gw_likelihood, em_likelihood)
priors_emgw = em_priors.copy()
priors_emgw.update(gw_priors)
\end{lstlisting}
Here, the first line sets up a joint likelihood (the product of the two individual likelihoods) and the functional interface for the code to interact correctly. The second line does the same, setting up a \prior{} object, automatically handling parameters that are shared. 

\begin{figure}
\centering
\includegraphics[width=0.49\columnwidth]{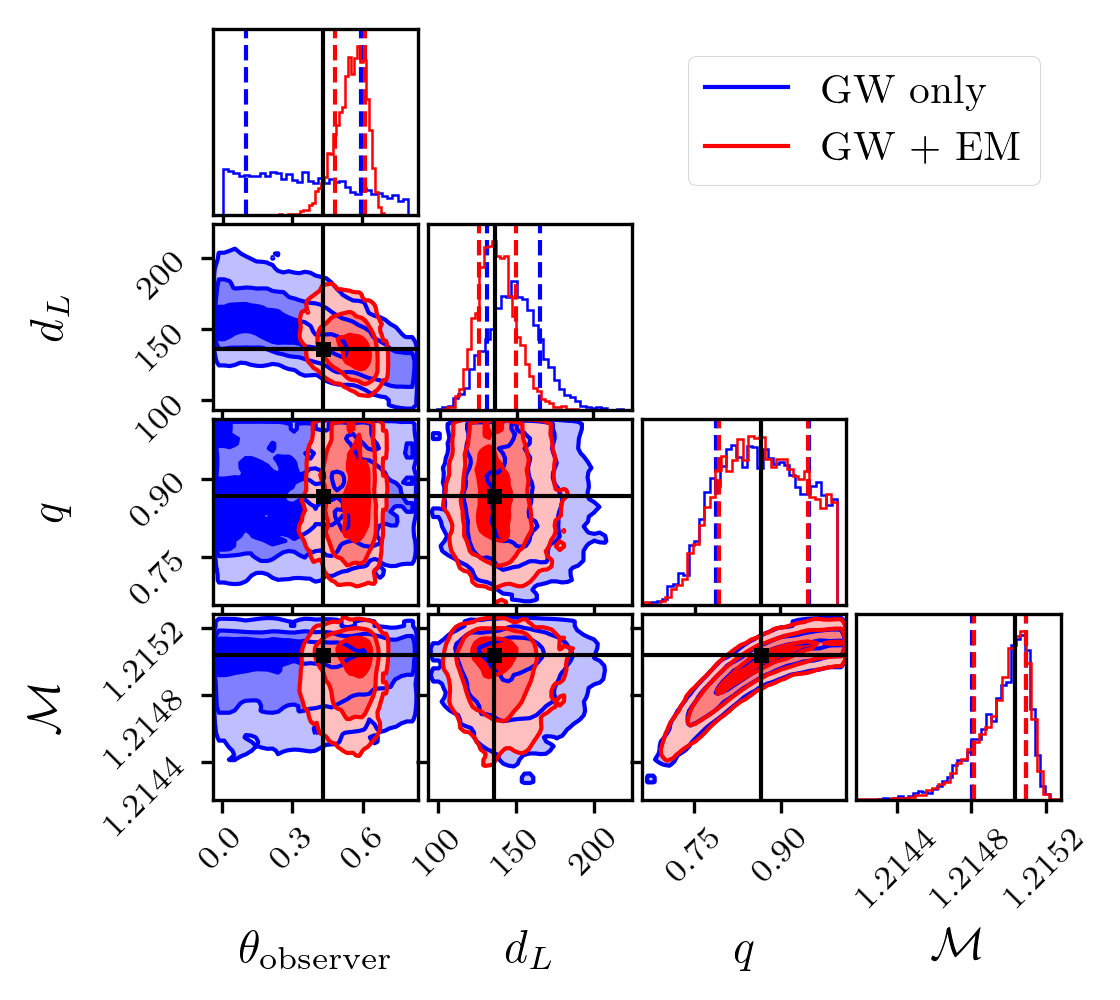} 
\includegraphics[width=0.49\columnwidth]{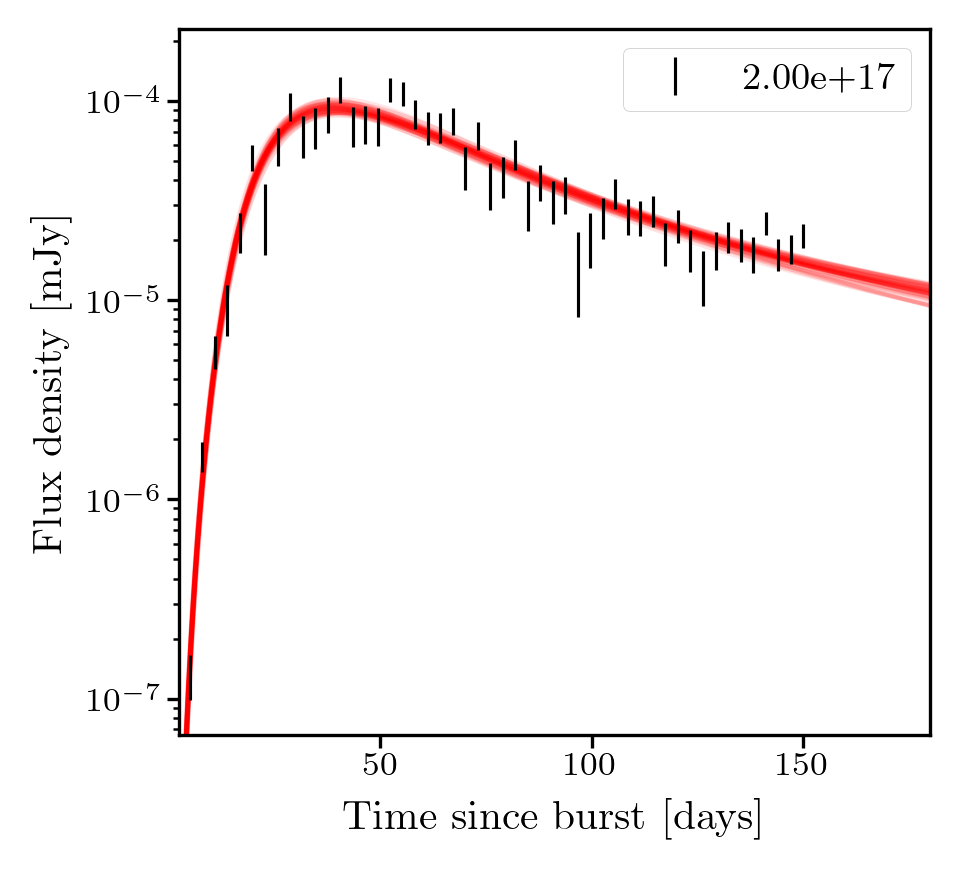}  
\caption{(Left) Corner plot showing the 1-3 $\sigma$ posterior on a subset of parameters with a GW only analysis (blue) and a GW + Afterglow analysis (red), with black lines indicating the input values of the simulation. The right-hand panel shows the lightcurve fit from the joint analysis.}
\label{fig:mmplot}
\end{figure}

Parameter estimation with the joint likelihood can then be performed via the \bilby{} interface, 
\begin{lstlisting}
result = bilby.run_sampler(joint_likelihood, priors=priors_emgw, label="emgw", outdir="joint")
\end{lstlisting}

In Fig.~\ref{fig:mmplot}, we show the constraints on various parameters provided by the above analysis, alongside constraints provided under the assumption that they are separate events. The orange lines indicate the true value of the simulation, indicating that the parameters are recovered correctly. In the right-hand panel, we show the fit to the simulated X-ray afterglow plotted via the \texttt{analysis} module. 
As expected, the primary benefit of including the afterglow is to break the distance-inclination angle degeneracy, clearly improving the estimate of distance and viewing angle for this hypothetical event. 

\section{Examples}\label{sec:examples}
We now go through a series of more general examples that demonstrate how \redback{} can be used to fit and infer properties of a variety of electromagnetic transients. We note that each of these examples are available as standalone scripts at \redbackurl{}. To aid readability of these examples in this paper, we avoid code snippets that are identical to the snippets described above. 

\subsection{Broadband afterglow - GRB170817A}
\begin{figure}
\centering
\includegraphics[width=0.50\columnwidth]{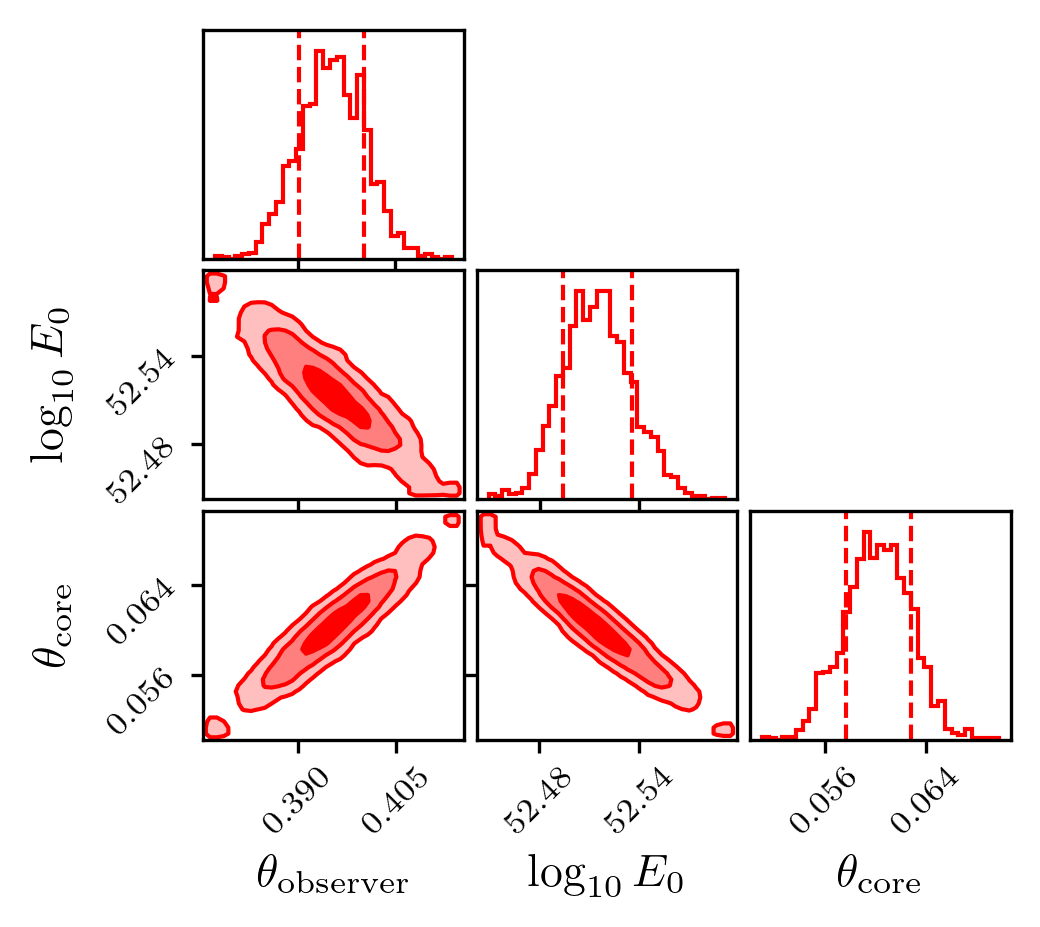} 
\includegraphics[width=0.49\columnwidth]{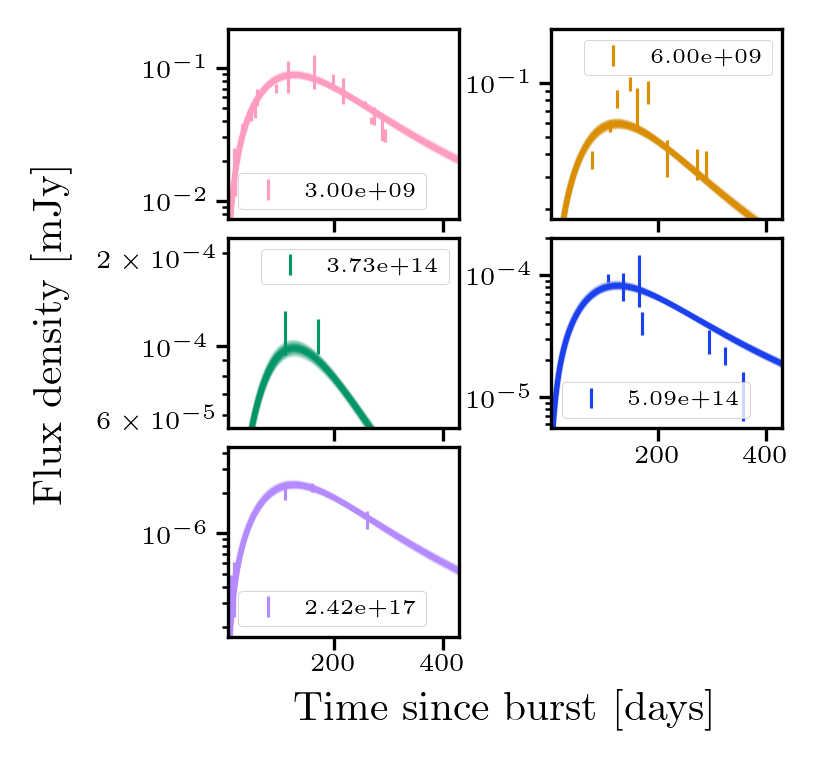}  
\caption{(Left) Posterior on the observers viewing angle, the isotropic equivalent energy of the afterglow and the opening angle of the relativistic jet from fitting the afterglow of GRB170817A with the different shading indicating the $1-3 \sigma$ credible intervals. (Right) Data of the afterglow of GRB170817A at multiple frequencies along with the lightcurves from a 100 random draws from the posterior.}
\label{fig:afterglowfigs}
\end{figure}

We first demonstrate how \redback{} can be used to fit private or simulated data by fitting the afterglow of GRB170817A~\citep{abbott17_gw170817_multimessenger, Hallinan2017, Alexander2018, Lamb2019_170817, Fong2019}. 
We must first load the data file and create an \texttt{afterglow} \transient{} object via the method described in Sec.~\ref{sec:creatingtransients}. 
After we have created the \transient{} object and have verified that the data looks correct (by plotting or by inspecting the \transient{} object), we are ready to fit. 
We know through many lines of evidence that GRB170817A was observed off-axis~\citeg{Fong2019, Alexander2018} and the jet was likely structured~\citeg{Lamb2019_170817, Fong2019}. 
Furthermore, many previous analyses have already fit the observations of GRB170817A to remarkable success. In particular, we can fit this data with a \texttt{gaussiancore} structured jet model from \texttt{afterglowpy}. As this model is already implemented in \redback{}, we simply need to specify this model as a string and load the associated prior. 
\begin{lstlisting}
model = "gaussiancore"
priors = redback.priors.get_priors(model=model)
\end{lstlisting}
These lines construct a \prior{} object using the default prior implemented in \redback{} for the \texttt{gaussiancore} model. To reduce inference wall-time, we can also fix some of the parameters of the model with values consistent as those found by~\citet{Ryan2020}. This can be done via, 
\begin{lstlisting}
priors["redshift"] = 1e-2
priors["logn0"] = -2.6
priors["p"] = 2.16
priors["logepse"] = -1.25
priors["logepsb"] = -3.8
priors["ksin"] = 1.
\end{lstlisting}
We note that we could have instead set a narrow Gaussian prior around these values instead of fixing these parameters. With these few lines, we are now almost ready for inference. As mentioned in Sec.~\ref{sec:overview}, several \redback{} models require additional keyword arguments; such as the frequencies at which each data point was was observed and the output format of the model (which must be the same as the data).
\begin{lstlisting}
model_kwargs = dict(frequency=afterglow.filtered_frequencies, output_format="flux_density")
\end{lstlisting}
Here, we have set up a model dictionary which contains the frequency of the data points (this can be easily extracted from the \transient{} object via the \texttt{filtered\_frequencies} attribute) and set the output format as flux density. 
We are now ready to fit via, 
\begin{lstlisting}
result = redback.fit_model(transient=afterglow, model="gaussiancore", sampler="dynesty", 
                           model_kwargs=model_kwargs, prior=priors, nlive=2000, resume=True)
\end{lstlisting}
Here we call the \redback{} \texttt{fit\_model} function, which takes as input the {\tt afterglow} object being fit, the name of the model, sampler, the prior, the model keyword arguments, and any other keyword arguments; and returns the \redback{} \result{} object. Here we have specified the sampler to be \dynesty{} via a string, but this could be any other sampler implemented in \bilby{}. We also specify some sampler settings such as the number of live points and the option to resume from a previous run. When finished, this will return the \redback{} \result{} object, which can be used to create a plot of the corner and a multi-band lightcurve to verify the fit via, 
\begin{lstlisting}
result.plot_corner(parameters=["thv", "loge0", "thc"])
result.plot_multiband_lightcurve(random_models=100)
\end{lstlisting}
Here, in the first line we have also passed a list of the parameters we wish to show and in the second asked for 100 randomly sampled lightcurves from the posterior to be plotted. Note that several other arguments can be passed into these functions to change aesthetics or the type of information displayed. These two plots are shown in Fig.~\ref{fig:afterglowfigs}. 

\subsection{Kilonova - AT2017gfo}
We now demonstrate how \redback{} can be used to fit a kilonova, in particular the kilonova that accompanied GW170817, AT2017gfo~\citep{abbott17_gw170817_multimessenger, Villar2017}.
For simplicity, we will fit a \texttt{one\_component\_kilonova\_model} implemented within \redback{} to observations of AT2017gfo~\citep{Villar2017}. Such a model is known to not provide a great fit to the data so this is merely a demonstration of \redback{} functionality. As mentioned in Sec.~\ref{sec:overview}, significantly more complex kilonovae models are available in \redback{} which have been previously shown to well explain the observations~\citeg{Villar2017, Bulla2019, Nicholl2021}.

\begin{figure}
\centering
\includegraphics[width=0.49\columnwidth]{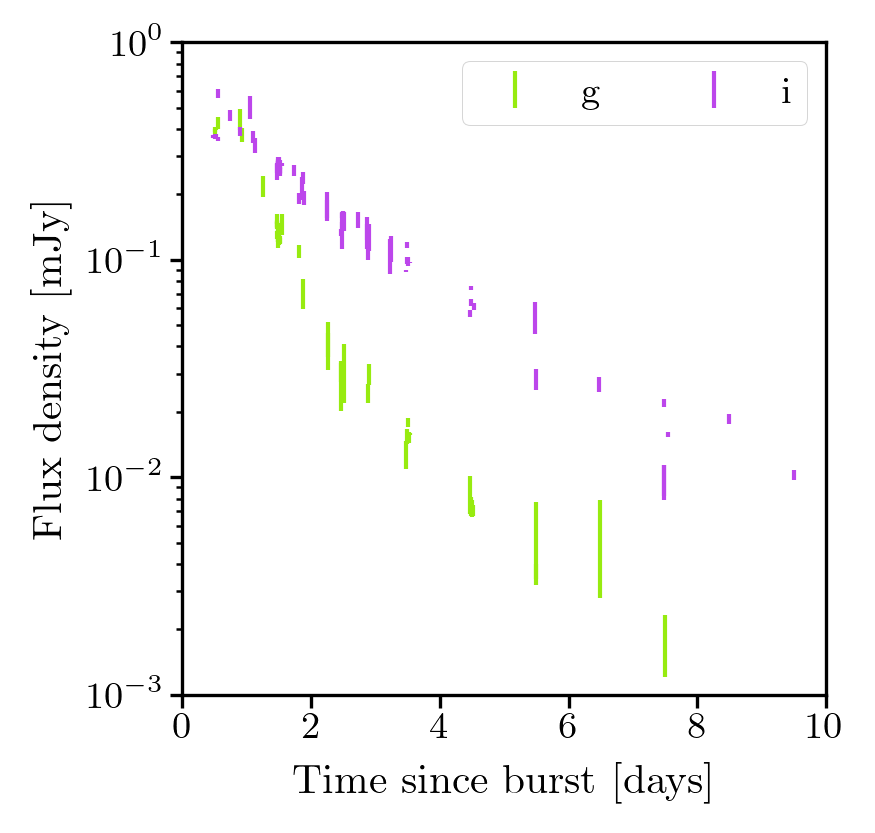} 
\includegraphics[width=0.49\columnwidth]{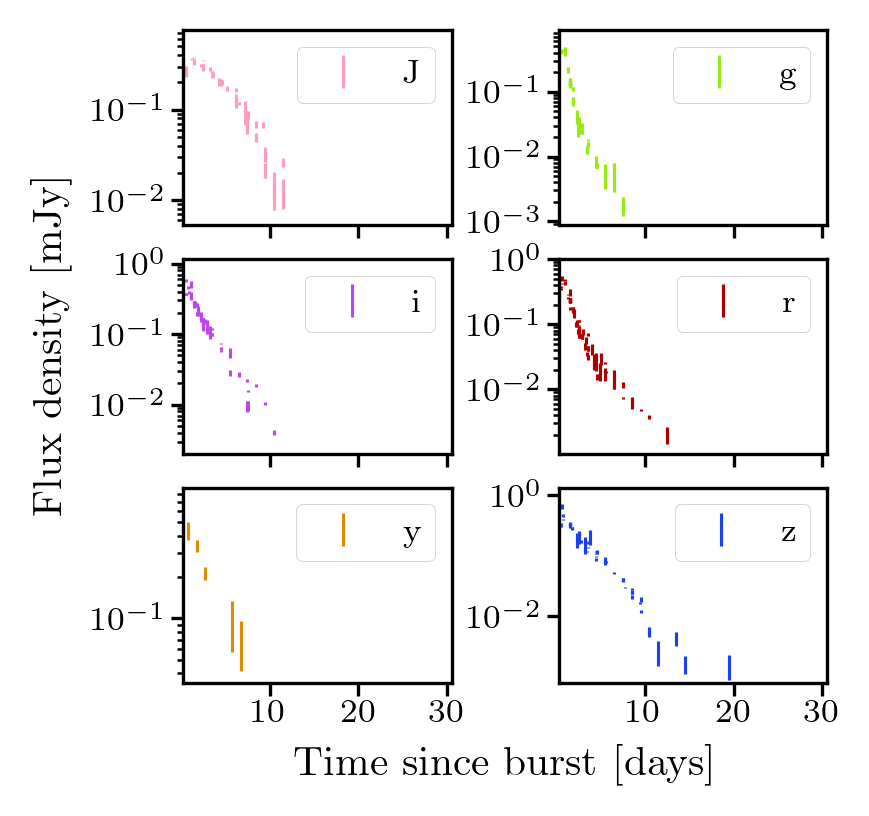}  
\caption{(Left) Data of AT2017gfo plotted through the \texttt{plot\_data} method. (Right) Data of the AT2017gfo plotted through the \texttt{plot\_multiband} method.}
\label{fig:kilonovafigs}
\end{figure}

The data of AT2017gfo is available at OAC~\citep{Guillochon2017}, which can be obtained via the code shown in Sec.~\ref{sec:gettingdata}. 
\begin{lstlisting}
data = redback.get_data.get_kilonova_data_from_open_transient_catalog_data(transient="at2017gfo")   
\end{lstlisting}
The above code calls the \getdata{} module to obtain the data for AT2017gfo from the OAC. As mentioned above, this will return a \texttt{pandas} data frame while also saving the data to disk. Users can manipulate the data as they would any other \texttt{pandas} object. However, for our purpose it is more useful to use this data to create an {\tt kilonova} object. This is done via
\begin{lstlisting}
kilonova = redback.kilonova.Kilonova.from_open_access_catalogue(
                    name="at2017gfo", data_mode="flux_density", active_bands=np.array(["g", "i"]))
\end{lstlisting}
Here we have created a \texttt{kilonova} \transient{} object, specifying the data mode to be flux density. We have also set the `g' and `i' bands as active, which will disable all other bands and only fit the active bands. This can be done to both reduce the computational time of inference but also for cases when the data or model are unreliable for specific filters. 
To ensure the data is correctly processed, we can plot the data via
\begin{lstlisting}
kilonova.plot_data(show=True, save=False, plot_others=False, band_colors={"g":"green", "i":"indigo"}, xlim_high=10)
fig, axes = plt.subplots(3, 2, sharex=True, sharey=True, figsize=(12, 8))
kilonova.plot_multiband(figure=fig, axes=axes, filters=["g", "r", "i", "z", "y", "J"])   
\end{lstlisting}
Here the first line will plot all the data onto one figure, where we have also passed additional arguments such a dictionary of the colors for each band, whether to plot the inactive bands, to not save and to show the plot and the upper limit on the x axis. Note that \redback{} returns the {\tt matplotlib} axes so several other plotting related things can be changed by the user directly or by passing in an additional keyword argument. 
The second line, will make a plot with one band per axes and we have also specified the specific filters we wish to display. Note that this functionality allows us to show data for a filter or fits for a filter even if that filter was set as inactive. Both figures are shown in Figure~\ref{fig:kilonovafigs}.

With the \transient{} object created and data verified through a plot, we are now ready to fit. As mentioned above, we will fit with the a one component kilonova model. However, we will now also demonstrate how a user can fit the data with a different likelihood and sampler. We skip steps to load a prior and set up model keyword argument dictionary as they are identical to the afterglow example above. 
\begin{lstlisting}
prior["sigma"] = Uniform(0.01, 0.0001, name="sigma", latex_label="$\sigma$")
function = all_models_dict[model]
sampler = "nestle"
\end{lstlisting}
Here, we first define a new prior on a parameter \texttt{sigma}, which is an additional parameter to be fit for, then use a convenience dictionary to get the \redback{} function for a one component kilonova model and specify the sampler to be used in inference as the \nestle{} sampler.
We note that \texttt{sigma} is the uncertainty in the typical Gaussian likelihood (i.e., $\sigma$), and if a user provides a prior but uses the standard (default) likelihood, this will overwrite the specific measured errors for a constant $\sigma$ that is estimated by sampling. However, here, we wish to demonstrate the use of a custom likelihood (either something provided by the user or a different likelihood already implemented in \redback{}), we can do this using the processed attributes from the \transient{} object via, 
\begin{lstlisting}
likelihood = redback.likelihoods.GaussianLikelihoodQuadratureNoise(x=kilonova.x[kilonova.filtered_indices], y=kilonova.y[kilonova.filtered_indices], sigma_i=kilonova.y_err[kilonova.filtered_indices],  function=function)
\end{lstlisting}
Here, we use a Gaussian Likelihood with an additional noise source, \texttt{$\sigma$} added in quadrature (that is fitted for) to the measured y errors. This likelihood is already implemented in \redback{} but a user could easily replace this likelihood with their own class. Then, users can use this likelihood in the fit via, 
\begin{lstlisting}
result = redback.fit_model(transient=kilonova, model=model, likelihood=likelihood, sampler=sampler, 
model_kwargs=model_kwargs, prior=priors)
\end{lstlisting}
With this simple change we can fundamentally change what we believe to be the data generation process and ensured that advanced users can easily change the likelihood and settings of the sampler, without ever digging into the \redback{} source code. 

\subsection{Supernova - SN1998bw}
\redback{} can also be used to fit supernovae. Here we fit the \texttt{arnett} model~\citep{Arnett1980, Arnett1982} implemented within \redback{} to observations of SN1998bw~\citep{Galama1998}.
We can acquire the data for SN1998bw through the OAC and API shown above and create a {\tt supernova} object. 

After ensuring that the data is obtained correctly we can set up the fit in a few lines of code. As the {\tt arnett} model is already implemented in \redback{} we can simply load up the default prior for this model via, 
\begin{lstlisting}
priors = redback.priors.get_priors(model="arnett")
priors["redshift"] = 0.0085
\end{lstlisting}
Here we have also fixed the redshift to the known redshift of SN1998bw. We can now set up the fit in another two lines of code. 
\begin{lstlisting}
model_kwargs = dict(frequency=supernova.filtered_frequencies, output_format="flux_density") 
result = redback.fit_model(transient=supernova, model="arnett", sampler="dynesty", model_kwargs=model_kwargs, 
prior=priors, nlive=500, clean=True, npool=4)
\end{lstlisting}
Here, we have also specified \texttt{npool=4} which will set up the \dynesty{} sampler with multiprocessing over 4 cores to reduce the wall-time of the analysis. We have also set the option \texttt{clean} to \texttt{True}, which ensures that \redback{} will restart this analysis from scratch and not resume from a previous analysis. 

\begin{figure}
    \centering
    \includegraphics[width=0.7\columnwidth]{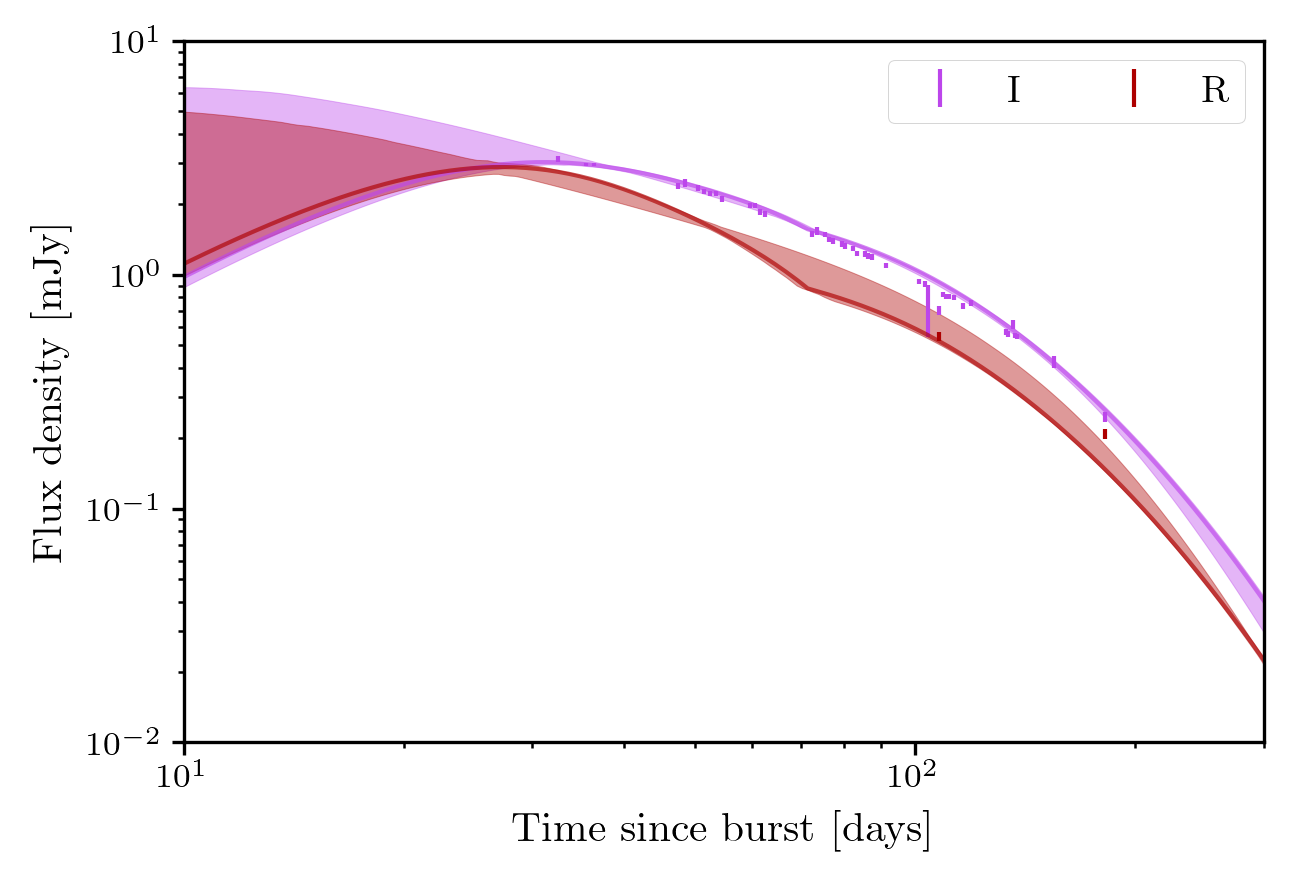}
    \caption{R and I band observations of SN1998bw alongside the $68\%$ credible interval from our fit.}
    \label{fig:sn1998bw}
\end{figure}

As with all other analysis, the fit returns a \redback{} \result{} object, which we can use to obtain posteriors on various parameters, or for plotting. For example, we can plot the lightcurve with the fit shown as a $68\%$ credible interval (shown in Fig.~\ref{fig:sn1998bw}) via, 
\begin{lstlisting}
ax = result.plot_lightcurve(uncertainty_mode="credible_intervals", plot_others=False, show=False, 
credible_interval_level=0.68)
ax.set_xscale("log")
ax.set_xlim(10, 300)
plt.show()
\end{lstlisting}
Here we have also returned the {\tt matplotlib} axes and used this to modify the xscale and xlimits of the plot. The fit demonstrates the large uncertainty at early times where there are no observations in these bands. 

\subsection{Tidal disruption events - PS18kh}
\begin{figure}
    \centering
    \includegraphics[width=0.95\columnwidth]{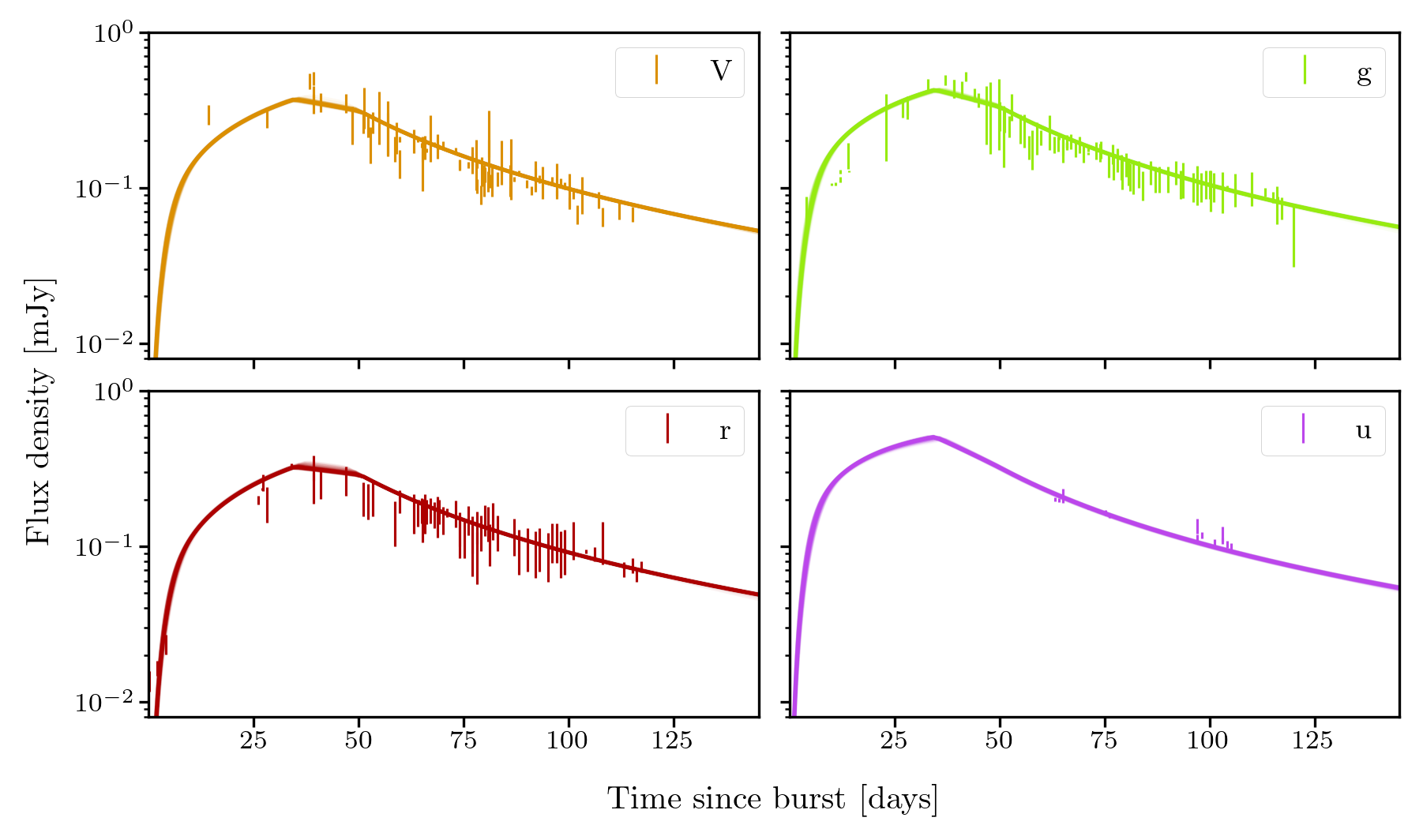}
    \caption{Multiband light curve of PS18bh along with the fitted lightcurve from a 100 random realisations randomly drawn from the prior.}
    \label{fig:tdemultiband}
\end{figure}

Here, we fit the \texttt{tde\_analytical} model implemented within \redback{} to multiband observations of, tidal disruption event, PS18kh~\citep{Holoien2019}. 
We acquire the data from OAC and create a {\tt tde} \transient{} object. We set only a subset of bands as active via, 
\begin{lstlisting}
tidal_disruption_event.active_bands = ["V", "g", "r"]
\end{lstlisting}
The rest of the code to fit is exactly like the afterglow example above. We can visualise our fit and make the predicted lightcurve (shown in Fig.~\ref{fig:tdemultiband}) for multiple filters, including a filter that we did not fit, e.g., the u-band, via 
\begin{lstlisting}
result.plot_multiband_lightcurve(random_models=100, filters=["V", "g", "r", "u"])
\end{lstlisting}
This is a useful verification exercise to understand which filters are driving the fit and whether the fits without a certain band are consistent with those observations. 

\subsection{X-ray afterglow of GRB070809 - millisecond magnetars}\label{example:xrayafterglow}
We now use \redback{} on an integrated flux or luminosity data by fitting the X-ray afterglows of a gamma-ray burst, by fitting the \texttt{evolving\_magnetar} model~\citep{SasmazMus2019} to \swift{} observations of GRB070809, specifically the integrated flux obtained from \swift{}-XRT. 

We acquire the BAT and XRT data of GRB070809 from \swift{} via the \getdata{} module
\begin{lstlisting}
redback.get_data.get_bat_xrt_afterglow_data_from_swift(grb="070809", data_mode="flux")
\end{lstlisting}
We construct an \texttt{afterglow} class instance via 
\begin{lstlisting}
afterglow = redback.afterglow.SGRB.from_swift_grb(name="070809", data_mode="flux",
                                                  truncate=True, truncate_method="prompt_time_error")
afterglow.analytical_flux_to_luminosity()
ax = afterglow.plot_data()
\end{lstlisting}
Here, we have specified to load the \texttt{flux} data for GRB070809 from \swift{}. This data typically also includes BAT data from the prompt phase which we do not wish to fit here. We truncate this data using the \texttt{prompt\_time\_error} method. 

\begin{figure}
    \centering
    \includegraphics[width=0.95\columnwidth]{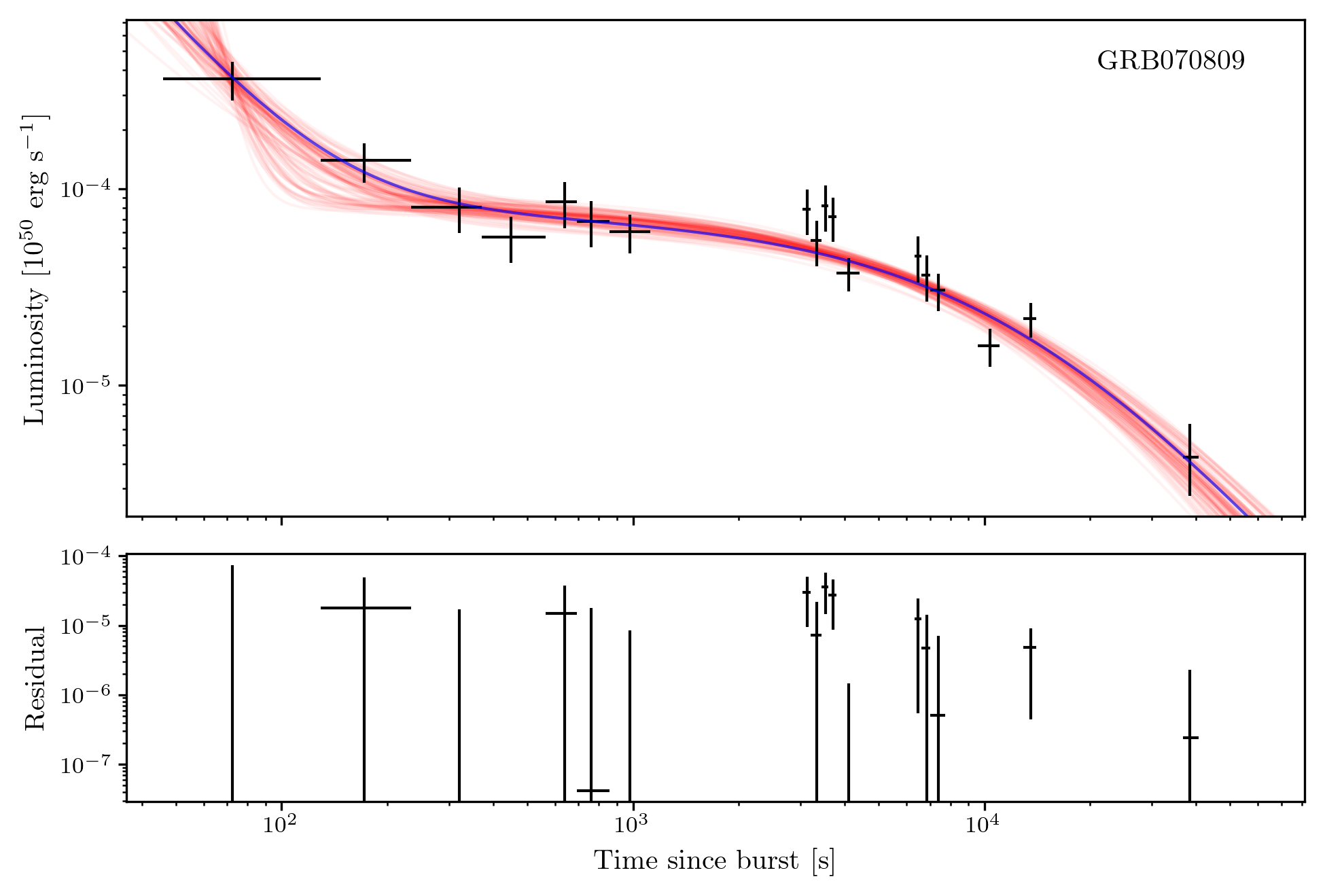}
    \caption{Residual plot obtained using \texttt{plot\_residual} method of the \result{} object. Here the top panel shows the data in black with maximum likelihood and 100 randomly drawn lightcurves in blue and red respectively with the bottom panel showing the residual between the data and the maximum likelihood model.}
    \label{fig:grb070809}
\end{figure}

The \texttt{evolving\_magnetar} model works on luminosity data. We could have provided this data when creating the {\tt afterglow} object but we also provide two convenience functions to generate this data, an analytical method which uses the GRBs photon index and a numerical method from {\tt Sherpa} which uses the spectrum. All details necessary for either method are obtained internally by \redback{} from the \swift{} Data Centre. 
Here, we use the analytical method to convert the integrated flux data to a luminosity. 
\begin{lstlisting}
afterglow.analytical_flux_to_luminosity()
\end{lstlisting}
Note, that this will automatically change the {\tt afterglow} objects data mode to luminosity. Beyond this point, the fitting workflow is identical to fitting any other transient i.e., 
\begin{lstlisting}
priors = redback.priors.get_priors(model="evolving_magnetar")
result = redback.fit_model(model="evolving_magnetar", sampler="dynesty", nlive=200, transient=afterglow,
                           prior=priors, sample="rslice", resume=True)
\end{lstlisting}
The above code first constructs a \prior{} object, using the default prior implemented in \redback{} for the \texttt{evolving\_magnetar} model. This is followed by code calling \texttt{fit\_model}. Note that here we do not need a dictionary for the model keywords as this model does not require any. 
We are again returned the \redback{} \result{} object which can be used to plot a corner plot, the lightcurve or obtain any other diagnostic about the inference/posterior. For these data modes however, it can be especially informative to show a plot of the lightcurve with the residuals. 
This can be obtained using \texttt{plot\_residual} method of the \result{} object. 
This generates Fig.~\ref{fig:grb070809}, where the top panel shows the data in black with maximum likelihood and random draws in blue and red respectively with the bottom panel showing the residual between the data and the maximum likelihood model. 
\begin{lstlisting}
result.plot_residual()
\end{lstlisting}

\subsection{Phase and Attenuation - SN2018ibb}
In previous examples, we have ignored two important aspects of fitting transients; 1) We often do not know when the explosion occured and 2) there is attenuation in the form of dust extinction from the host galaxy. 

In this example, we show how to fit data while measuring the unknown explosion time and including extinction. We will also demonstrate how to fit in magnitudes and adding a new filter to \redback{} and {\tt SNCosmo}. We will do this by fitting a supernova, in particular, the UV-to-NIR light curve of the superluminous supernova SN~2018ibb~\citep{Schulze2023}.

As previous examples, we can load the private data for SN~2018ibb and create a {\tt Supernova} \transient{} object via

\noindent First, we read in the private data.
\begin{lstlisting}
import pandas as pd
data = pd.read_csv("SN2018ibb_photcat_Redback.ascii", sep="")
sn=redback.transient.Supernova(name = "SN2018ibb",
data_mode = "magnitude", time_mjd = data["MJD"].values,
magnitude = data["MAG"].values, magnitude_err = data["MAG_ERR"].values, bands = data["band"].values,
use_phase_model = True)
\end{lstlisting}

In contrast to the previous examples, we fit the data in magnitude space. Furthermore, we set \verb|use_phase_model = True| because we do not know the explosion date. We also specify time values in MJD instead of days since explosion. When fitting a model to such data, a user must then add a prior on the explosion time which will then be sampled over. We note that \verb|use_phase_model = True|, will also change plotting labels to account for the change. 

Before we can fit the magnitude data of SN~2018ibb, we must first ensure that all filters of the observations are available in \redback{}. We note that this is only a concern when fitting photometry in magnitudes or flux as this requires the full transmission curve of every filter rather than a reference wavelength. 
Some of the observations of SN~2018ibb were performed with the GROND camera mounted at the 2.2\,m MPG telescope. The GROND filters (in our example \verb|grond::i| and \verb|grond::z|) are not part of {\tt SNCosmo} distribution that is used internally within \redback{} for filter definitions. 
After retrieving the filters, for instance, from the Spanish Virtual Observatory\footnote{\href{http://svo2.cab.inta-csic.es/theory/fps/}{http://svo2.cab.inta-csic.es/theory/fps/}}~\citep{Rodrigo2012a}, we add them to {\tt SNCosmo} and by extension \redback{}, via
\begin{lstlisting}
from   astropy.io import ascii
import astropy.units as u
import sncosmo

filter_files = [
        "/PATH/WHERE/YOU/STORED/FILTER_CURVES/GROND_I.dat",
        "/PATH/WHERE/YOU/STORED/FILTER_CURVES/GROND_Z.dat",
        ]

filter_names = ["grond::i", "grond::z"]

for f, fname in zip(filter_names, filter_files):
    _data = ascii.read(fname)
    band  = sncosmo.Bandpass(_data["col1"], _data["col2"], name=f, wave_unit=u.angstrom)
    sncosmo.register(band, f, force=True)
\end{lstlisting}

We can set up the rest of the inference workflow, first set up the model and the prior via 

\begin{lstlisting}
model      = "t0_supernova_extinction"
base_model = "arnett"

priors     = redback.priors.get_priors(model=model)
priors.update(redback.priors.get_priors(model=base_model))
\end{lstlisting}

Here we choose the \verb|t0_supernova_extinction| model, which has the explosion time and magnitude of extinction as a free parameter. This model itself does not contain any physics and must be specified an additional physical model. For simplicity, we use the physical \verb|arnett| model as the base model. The last two lines of code just set up the prior object to include the parameters of both models.

We must now also set priors on the explosion time, the extinction magnitude and update the prior on the ejecta mass as SN~2018ibb requires an extraordinary amount of ejecta~\citep{Schulze2023}.

\begin{lstlisting}
from bilby.core.prior import Uniform

# Allow the explosion date to be up to 200 days before the first detection
priors["t0"] = Uniform(minimum = data["MJD"].values.min()-200, maximum = data["MJD"].values.min()-1, name = "t0", latex_label = r"$t_{\rm expl.}$")

priors["mej"] = Uniform(minimum = 1, maximum = 260,  name = "mej", latex_label = r"$M_{\rm{ej}}~(M_{\odot})$")
                                           
# Extinction
priors["av"]  = Uniform(minimum = 0, maximum = 1, name = "av",latex_label = r"$A_V$ (mag)")
\end{lstlisting}

\begin{figure}
    \centering
    \includegraphics[width=0.95\columnwidth]{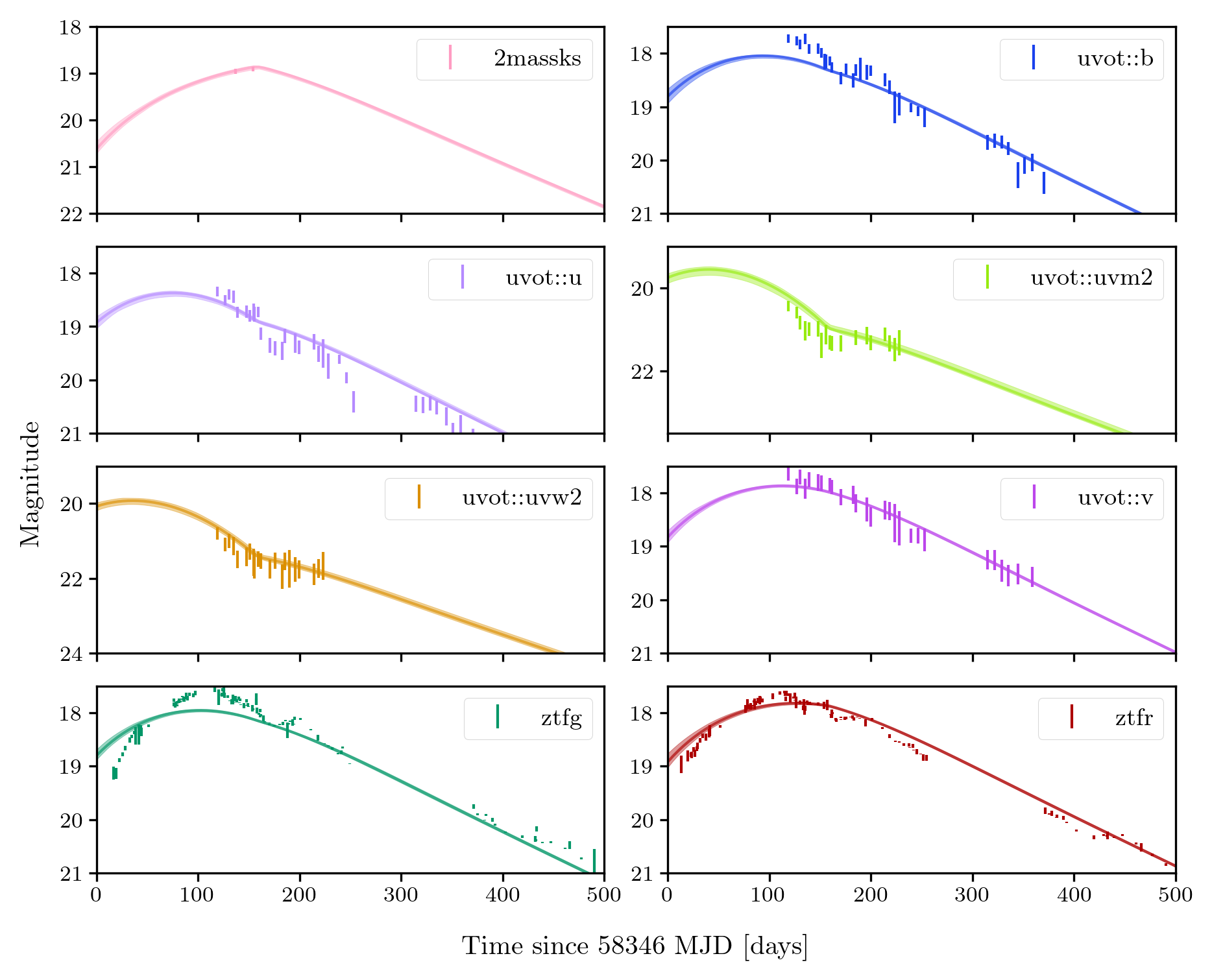}
    \caption{Multiband light curve of SN2018ibb along with the $68\%$ credible interval lightcurve fit from the posterior.}
    \label{fig:2018ibb}
\end{figure}

With the model specified and prior set up, we can now fit via, 
\begin{lstlisting}
model_kwargs = dict(bands = sn.filtered_sncosmo_bands, base_model = base_model, output_format = "magnitude")

result = redback.fit_model(transient = sn, model = model, model_kwargs = model_kwargs, 
prior = priors, plot = True)
\end{lstlisting}

We note that as we are fitting with a base model and in magnitudes, there are some minor differences to the \texttt{model\_kwargs}, namely that we must now specify a list of bands for the data points instead of frequency and must specify the base model. In the {\tt fit\_model} argument, we have also set {\tt plot=True}, which will automatically generate the fitted lightcurve after inference finishes. In Fig~\ref{fig:2018ibb}, we show the lightcurve fit generated with the above code and the {\tt result.plot\_multiband\_lightcurve()}.

\bsp    
\label{lastpage}
\end{document}